\documentclass[manuscript,iop]{emulateapj}
\usepackage{graphicx}
\usepackage{epstopdf}
\usepackage{epsf}
\usepackage{amsmath, amsthm, amssymb}

\def\la{\mathrel{\hbox{\rlap{\hbox{\lower4pt\hbox{$\sim$}}}\hbox{$<$}}}}
\def\gt{\mathrel{\hbox{$>$}}}

\def\ga{\mathrel{\hbox{\rlap{\hbox{\lower4pt\hbox{$\sim$}}}\hbox{$>$}}}}

\newcommand{\msun}{\mbox{${M}_{\odot}$}}
\newcommand{\mstar}{M_{\star}}

\shortauthors{Skelton et al.}
\shorttitle{Modeling the red sequence}

\begin{document}

\title{Modeling the red sequence: Hierarchical growth yet slow luminosity evolution}
\author{Rosalind E. Skelton,\altaffilmark{1} Eric F.\ Bell\altaffilmark{2} and Rachel S. Somerville \altaffilmark{3}}
\altaffiltext{1}{Department of Astronomy, Yale University, 260 Whitney Avenue, New Haven, CT 06511, USA}
\email{ros.skelton@yale.edu}
\altaffiltext{2}{Department of Astronomy, University of Michigan, 500 Church Street, Ann Arbor, MI 48109-1042, USA}
\altaffiltext{3}{Department of Physics and Astronomy, Rutgers, The State University of New Jersey, 136 Frelinghuysen Road, Piscataway, NJ 08854, USA}

\begin{abstract}

We explore the effects of mergers on the evolution
of massive early-type galaxies by modeling the evolution of their
stellar populations in a hierarchical context. We investigate how a
realistic red sequence population set up by $z\sim 1$ evolves under different assumptions for the merger and star
formation histories, comparing changes in color, luminosity, and
mass. The purely passive fading of existing red sequence galaxies,
with no further mergers or star formation, results in dramatic changes
at the bright end of the luminosity function and color--magnitude
relation. Without mergers there is too much evolution in luminosity at
a fixed space density compared to observations. The change in color
and magnitude at a fixed mass resembles that of a passively evolving
population that formed relatively recently, at $z\sim2$. Mergers
amongst the red sequence population (``dry mergers") occurring after
$z=1$ build up mass, counteracting the fading of the existing stellar
populations to give smaller changes in both color and luminosity for
massive galaxies. By allowing some galaxies to migrate from the blue
cloud onto the red sequence after $z=1$ through gas-rich mergers, younger
stellar populations are added to the red sequence. This
manifestation of the progenitor bias increases the scatter in age and results in even smaller changes in color and
luminosity between $z=1$ and $z=0$ at a fixed mass. The resultant
evolution appears much slower, resembling the passive evolution of a
population that formed at high redshift ($z\sim3 - 5$), and is in closer agreement with observations. We conclude that measurements
of the luminosity and color evolution alone are not sufficient to
distinguish between the purely passive evolution of an old population
and cosmologically motivated hierarchical growth, although these
scenarios have very different implications for the mass growth of
early-type galaxies over the last half of cosmic history.
\end{abstract}

\section{Introduction}
\label{sec:intro}

Galaxies in the local universe are found in a bimodal distribution in color--mass (or magnitude) space \citep{Strateva01,Kauffmann03a,Blanton03, Baldry04}. The so-called blue cloud is made up mostly of star-forming, disk galaxies, while the majority of quiescent galaxies with early-type morphologies lie on a tight relation between color and mass known as the red sequence. The tilt in the color--mass relation (CMR) of early-type galaxies can largely be attributed to an increase of metallicity with mass, with more metal rich galaxies having redder colors \citep{Kodama97, Gallazzi06}. The distinction between red sequence and blue cloud galaxies is particularly clear at $z<1$ \citep[e.g.,][]{Bell04, Borch06,Faber07,Brown07}, and using optical--NIR colors or mid-IR diagnostics the populations can be separated reasonably well out to $z\sim3$ \citep[][]{Labbe05, Williams09,Brammer09,Whitaker11}. 

Although most star formation occurs in blue cloud galaxies, the total mass in galaxies on the red sequence has grown by an order of magnitude since $z\sim2$ \citep{Arnouts07,Ilbert10, Brammer11} and by a factor of two since $z\sim1$ \citep[the epoch we focus on in this paper; e.g.,][]{Bell04, Faber07, Brown07}.  \citet{Bell07} showed that the increase in mass on the red sequence can largely be explained by an influx of blue cloud galaxies that have recently had their star formation shut down. \citet{Harker06} and \citet{Ruhland09} used stellar population synthesis models to show that the scatter in a red sequence that is built up through the quenching
of star formation in blue cloud galaxies is consistent with observations of the evolution of the CMR. These works did not make any assumptions on what shuts the star formation down. The merging of gas-rich disk-like galaxies is a likely mechanism for rapid morphological \citep{Toomre77, Schweizer92,Barnes96} and color transformation, with some kind of feedback process usually invoked to prevent further gas cooling and star formation \citep[e.g., active galactic nucleus (AGN) ``radio mode" feedback;][]{Croton06, Bower06, Somerville08}. Most massive early-type galaxies are spheroidal in shape suggesting that they have formed through major mergers \citep{vanderWel09}. Models in which spheroids form in gas-rich mergers, with a period of related quasar activity, have been fairly successful in matching a number of observations \citep{Benson03, Croton06, Bower06, Somerville08, Hopkins08b, Hopkins09b}. 

We maintain, and will discuss below, that the growth in mass on the red sequence has been well measured only for galaxies near the knee of the luminosity function (LF; $\sim~L^*$); it is still unclear whether the massive end of the red sequence has seen significant growth since $z=1$. If such growth occurs, it is likely to take place through the merging of gas-poor galaxies that already lie on the red sequence since there are relatively few luminous blue galaxies from which the most massive early-type galaxies could form \citep{Blanton06}. The role of mergers in the build up of massive early-type galaxies, particularly over the epoch from $z=1$ to $z=0$, is strongly debated. In this paper we look at how a red sequence in place by $z \sim 1$ evolves with and without further mergers and star formation. We show that observations of color and luminosity evolution that have been taken as support for a limited role for mergers in the past are consistent with the changes we find using hierarchical models.

The outline of the paper is as follows: Section~\ref{sec:bg} discusses previous work on the evolution of early-type galaxies, the support for a hierarchical picture of early-type evolution, and why early-type galaxies were thought to evolve passively from high redshift. Section~\ref{sec:models} describes
the model set up, building on the simple toy model presented in
\citet{Skelton09a}. In Section~\ref{sec:results} we present the resultant
evolution of the color, luminosity and mass distributions in the
models and compare to observations. We consider the implications for
massive galaxies in particular and compare the changes in color and
magnitude to those of an ancient simple stellar population. In
Section~\ref{sec:discussion} we discuss how the various model choices
affect our results and compare to previous work.  The conclusions that
can be drawn on early-type evolution are summarized in
Section~\ref{sec:summary}. We adopt a cosmology with $\Omega_M = 0.3$,
$\Omega_\Lambda = 0 .7$ and $H_0 = 100 h$~km~s$^{-1}$~Mpc$^{-1}$ with
$h = 0.7$ and use Vega magnitudes throughout.

\subsection{Background}
\label{sec:bg}

Gas-poor or ``dry'' mergers have been observed both in the local universe \citep{vanDokkum05,McIntosh08,Tal09} and at higher redshifts \citep{Bell06a, Lotz06, Lin08, Williams11}, and a non-negligible fraction of massive galaxies are found to occur in close pairs that are likely
to merge \citep[][although see \citealt{Masjedi06}, who measure very low merger rates for luminous red galaxies, LRGs]{Bell06b, Robaina10, Williams11}. Recent measurements place the fraction of $M_* \gt 5 \times 10^{10}{{M_{\sun}}}$ galaxies in close pairs at 1 to 3\% out to $z=1.2$, implying that galaxies with masses of $\gt 10^{11}{{M_{\sun}}}$ have undergone 0.7 mergers on average since this time \citep[][]{Robaina10}. This merger fraction seems to be sufficient to explain the growth of the red sequence at the massive end since $z=1$ \citep{Robaina10, Man11}. A particularly large uncertainty in interpreting measurements of the merger fraction as a growth rate is
the timescale over which mergers can be identified. The timescale will be different for each measurement technique (e.g., visual or automated morphological selection, close pairs) and is dependent on the properties of the progenitor galaxies and the orbit \citep{Lotz10a, Lotz10b}. Recent work that calibrates the observability timescale for each method using high-resolution simulations in combination with cosmological models has shown a reassuring consistency between a number of different measurements of the major merger rate and model predictions \citep{Lotz11}. 

Less direct approaches for estimating the extent to which dry mergers build up massive red galaxies, such as measuring their clustering and size evolution, support a degree of merging activity since $z=1$ but are not yet conclusive. 
Massive early-type galaxies have evolved substantially in size over this period (a factor of
two growth for a given mass, \citealt{vanderWel08}). This can be
accounted for by the increase in size of individual red galaxies
through dry mergers, combined with an increase in the number density
of the red sequence galaxy population over time
\citep{vanderWel09b}.  Minor mergers may be particularly important for the size evolution of early-type galaxies, as well as a viable means of explaining the evolution of the velocity dispersion function since $z\sim1$ \citep{Bezanson09,vanderWel09b,Naab09, Hopkins09c, Hopkins10c}.  Clustering analyses using LRGs in the Sloan Digital Sky Survey (SDSS) and 2SLAQ surveys \citep{Wake08} and the Bo\"{o}tes field \citep{White07} showed little evolution in the clustering amplitude with redshift. Interpreting these results in the light of the halo model suggests that LRGs do not evolve purely passively but that a significant fraction of satellite galaxies (30\%--50\%) must either merge with the central LRG or be disrupted into the intracluster light \citep{Conroy07}. \citet{Tojeiro10} also examine the clustering evolution of LRGs and find evidence for luminosity growth that is inconsistent with passive evolution. The growth appears to occur preferentially at the faint end of their galaxy selection, supporting an influx of galaxies into the sample through mergers \citep[see also][]{Tojeiro11b}. 

To determine how the massive end of the red sequence evolves, one would ideally want to measure the mass function (MF); however, this requires combining the measured LF with knowledge of the mass-to-light ratio ($M/L$), which also evolves. Stellar population synthesis models are usually used to estimate $M/L$, requiring assumptions on the age and star formation histories of the galaxies. To measure these quantities, the evolution must be known --- this is one of the reasons that passive evolution (the evolution of a simple stellar population formed within a short time at high redshift, with a single metallicity) is often assumed for early-type galaxies. 

There is still large uncertainty in both the measurement of the number density evolution and its interpretation. If one corrects the observed LF for the evolution expected for a purely passively fading old population, there is very little remaining evolution at the bright end  since $z\sim1$ \citep{Cimatti06, Wake06, Brown07, Cool08, Banerji10}. These results have been used to infer that the number density of massive red galaxies has not changed substantially over the last 8 billion years and that much of the mass is already in place by this time.\footnote{Perhaps contrary to expectation, one of the largest difficulties in this analysis is determining the bright end of the LF at low redshift to high precision. Bright galaxies in the local universe have extended outer envelopes (possibly caused by dry merging) making it difficult to estimate the total magnitude and accurately subtract the background \citep[e.g.,][]{Bernardi07a, Lauer07}. Small photometric errors translate into a large error in number density because of the exponential drop of the LF toward the bright end.} If the early-type galaxy population is continuously supplemented by quenched blue cloud galaxies, the assumption of slow luminosity evolution made in the interpretation the evolution of the LF will be incorrect. The build up of stellar mass through mergers may act to compensate faster luminosity and color evolution, resulting in changes that look very much like passive evolution. This is related to the so-called ``progenitor bias'' \citep{vanDokkumFranx01}; galaxies that have recently moved onto the red sequence through morphological transformation will be missing from early-type galaxy samples at high redshift, so that at early times, any sample of early-type galaxies consists of only the oldest progenitors of today's early-type galaxies. Converting the LF of red galaxies measured by \citet{Brown07} to a MF using the $M/L$ evolution from Fundamental Plane measurements \citep{vDvdMarel07} rather than using stellar population models gives rise to stronger evolution \citep{Robaina10}. Other authors have recently found a similar increase in number density (approximately a factor of four) at the massive end using the galaxy MF \citep{Matsuoka10}. A recent analysis that uses clustering measurements to select likely progenitors of early-type galaxies at high redshift measures a factor of $\sim5$ growth in number density, supporting substantial growth through dry mergers \citep{Padilla11}. 

The uniformity of the stellar populations in early-type galaxies suggested by the tight CMR in clusters has long been used as an argument that they formed relatively quickly at high redshift \citep{Ellis97, BKT98, Andreon11}. However, recent work on the evolution of the red sequence has shown that it is difficult to match both the normalization and change in color with a purely passive, old model \citep{Kriek08, Ruhland09, Whitaker10}.  The small scatter in the colors of early-types in clusters was thought to impose a limit to the amount of growth that could occur via merging \citep{BKT98}. In \citet[SBS09 hereafter]{Skelton09a} we explored how the amount of dry merging expected in a standard hierarchical model affects the scatter and slope of the CMR. We assumed that major gas-rich mergers move galaxies from
the blue cloud onto the red sequence, placing them onto a ``creation
red sequence'' given by the local relation between color and
magnitude. We calculated where red galaxies at low redshift lie in
color-magnitude space by combining the fluxes of their progenitors,
assuming that merging of galaxies on the red sequence 
does not involve further star formation. By using the local CMR as a reference, the model essentially assumed that early-type galaxies are evolving passively and at the same rate. With this simple model we showed
that dry merging can reduce the scatter and flatten the bright end of
the CMR. The resulting relation was found to agree well with the
observed red sequence from the SDSS. 

In this paper we incorporate more realistic evolution of the stellar populations into the model described above to test how mergers affect the color, luminosity, and mass evolution of the red sequence since $z =1$.  Our aim is not to make a complete, accurate model of
the galaxy distribution; the point is rather to show how the
progenitor bias expected from hierarchical growth in a standard cold
dark matter cosmology affects the evolution of massive galaxies. The models, very much in the spirit of earlier work by
\citet{BKT98, vanDokkumFranx01,Blanton06,Skelton09a}, and others, use stellar population
synthesis modeling to follow the luminosity evolution of galaxies, but
set the model galaxies into a hierarchical context using merger trees extracted from a semi-analytic model (SAM) of galaxy formation. We assume that the
red sequence is built up mainly through major gas-rich mergers. With a
reasonably realistic red sequence population in place at $z =1$, we
explore the consequences of subsequent gas-poor and gas-rich merging. We compare this to the
evolution of the same population undergoing only passive evolution with no mergers
after $z=1$. We show that at fixed mass, dry merging and recent additions to the red sequence decrease the change in color and
luminosity. The resultant evolution is consistent with that of a
population that formed early and evolved passively, {\it even though
  there has been significant merging activity}.

\section{Model Framework}
\label{sec:models}

To investigate how the colors and magnitudes of galaxies {\it  evolve} in a hierarchical universe, we incorporate the evolution of
stellar populations of different metallicities into the model described in SBS09. Galaxy
merger trees from the \citet[S08 hereafter]{Somerville08} SAM provide the
hierarchical base for the models. Rather than using the full SAM, a
complex model that has successfully reproduced a number of
observations, we aim to disentangle the effects that mergers have on
early-type galaxy evolution with a simplified model setup. The SAM provides
cosmologically motivated merger statistics that have been shown to
match well with observations \citep{Bell06b, Jogee09, Lotz11}. 

In the S08 SAM the dark matter merger
trees are determined using the extended Press-Schechter formalism, as
described in \citet{Somerville08, Somerville01, SK99}. Analytic prescriptions for gas cooling (S08, Section 2.2), star formation (S08, Section 2.5), supernova feedback (S08, Section 2.7) and AGN feedback from winds (S08, Section 2.10) and radio jets (S08, Section 2.11) determine how galaxies form and evolve within the dark matter halos. A modified Chandresekhar dynamical friction formula that takes into account mass loss due to tidal stripping \citep{Boylan-Kolchin08} is used to calculate the time taken for galaxies to merge after the merger of their dark matter halos. Quenching occurs when the central black hole becomes large enough to prevent further gas from cooling through radio mode feedback. This often occurs for the first time immediately following a major merger, as these mergers trigger a phase of rapid accretion leading to significant growth of the black hole. Due to the self-regulation of black hole accretion, there is a correlation between the mass of the remnant central black hole and the mass of the bulge component of the galaxy, providing a natural link between quiescent galaxies and spheroid-dominated morphologies. At high redshifts and in low mass halos, further cold gas may be accreted via cold dense filaments that are impervious to heating by the radio jets, so star formation may begin again some time after a major wet merger has occurred, even if the existing gas is removed during the merger process. These physical processes provide the motivation for our simple model but are not included in any detail. 

We construct galaxy merger trees from the SAM output, using the halo and galaxy identity numbers to track galaxies as they merge, and record the stellar, cold gas and dark matter masses for the two progenitor galaxies involved in every merger. These are used as input to the toy model.  We infer the metallicity of each merging galaxy from its stellar mass using a simple approximation to the mass--metallicity
($M$--$Z$) relation given by 
\begin{equation}
 Z = -0.082 + 0.0098 \log\left(\frac{M_*}{{M_{\sun}}}\right).
\label{eqn:mz}
\end{equation}
This relation was obtained by fitting a straight line through the median metallicity as a function of stellar mass from \citet{Gallazzi05} and adjusting the normalization by 0.004 so that the color-magnitude relation of the model approximately matches the observations at z=0.9. Solar metallicity is 0.02. We assume that the $M$--$Z$ relation does not evolve and the
metallicity of each galaxy remains the same for the duration of its
star formation. Although these are clearly simplifications, they are
defensible choices in the light of this model framework.  The existence of a mass-gas metallicity relation has been established out to $z\sim3$ \citep{Erb06}, but it is very difficult to pin down how the relation evolves observationally and there is large uncertainty in the calibration of the observed relation \citep{Kewley08}. \citet{Erb06} find a factor of 2 offset between the relation at $z\sim2$ and $z\sim0$ but note that the uncertainty is at about the same level. The rate at which the relation evolves seems to be mass dependent, with very little evolution since $z\sim0.8$ for massive galaxies (\citealt{Savaglio05,Zahid11}; but see \citealt{Moustakas11}, who find lower metallicities by 30\%--60\% at $z=0.7$ with little mass dependence).

In the toy model, the red sequence forms from the remnants of major gas-rich mergers. We begin by identifying the major wet mergers occurring along every branch of each merger tree. Major mergers are defined to have a mass ratio between 1:1 and 1:4, where the total baryonic and dark matter mass within twice the Navarro--Frenk--White (NFW) scale radius \citep{Navarro97} is used to determine the ratio (see S08). The radius of interest corresponds to approximately 60~kpc for a Milky Way-size halo. To distinguish between wet and dry mergers, we assume a gas fraction threshold of 20\%, where the gas fraction is defined as the ratio of cold gas mass to total baryonic mass (cold gas and stars). This distinction is somewhat arbitrary but the region of color--magnitude space dominated by galaxies with lower gas fractions corresponds closely to the red sequence in the full S08 SAM. This threshold also produced the closest match between the model and observed red sequences in SBS09. If either of the galaxies has a gas fraction of more than 20\%, the merger is classified as wet.
 
\begin{figure*}[ht]
  \begin{center}
    \leavevmode
       \plotone{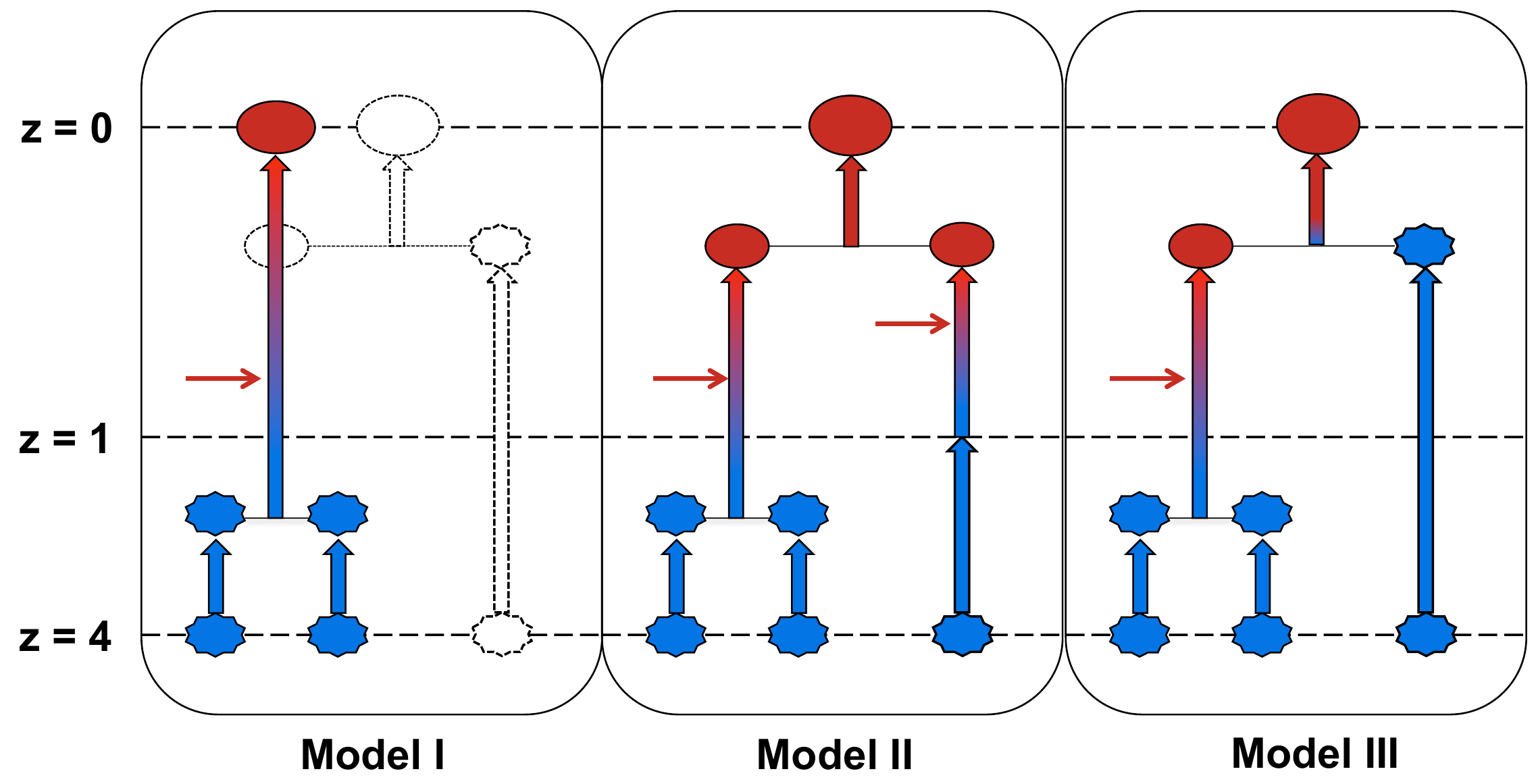}
	\caption[Schematic illustrating the evolution of merging galaxies in the three models]{Schematic illustrating the evolution of a galaxy that has undergone a major merger at high redshift and a second merger at $z < 1$ in the three models. Arrows with a color gradient indicate that the stellar populations are evolving passively. Continuing star formation is indicated by the blue arrows. Star formation stops at the time of the first major merger. In Model I, no mergers take place after $z=1$, as shown by the shaded path. In Model II, star formation is quenched at $z\sim1$ and galaxies that do not undergo a merger before $z=1$ start to fade passively at this time.  In Model III, such galaxies continue to form stars until they undergo a major wet merger. The small red arrows show approximately when the galaxies would be classified as red sequence galaxies. \label{fig:cartoon}} 
  \end{center}
\end{figure*}

We assign each galaxy involved in a major wet merger a star formation history, assuming that the star formation in both progenitor galaxies was truncated at the time of the merger. The masses and fluxes of the two progenitors are combined and evolved from the time of the merger to the time the remnant galaxy is ``observed". If further dry or minor mergers occur before the time of observation, the flux (mass) of the secondary galaxy are added to the flux (mass) of the main progenitor.  The secondary galaxy's properties are determined by following its merger tree back until a major wet merger occurred, and so on, recursively, for each branch of the merger tree. If no major wet merger occurs along one branch, star formation is assumed to have stopped at the first merger where the main progenitor is recorded as gas poor. Only mergers where the mass ratio is between 1:1 and 1:10 are considered.

In keeping with the simplicity of the model, we choose a constant star
formation history for each galaxy from some formation redshift $z_f$ up until the time
at which star formation stops, with the addition of a burst of star formation at the time of the merger. We present models with $z_f = 4$
below, but find that the exact choice does not strongly affect the
results (see Section~\ref{sec:discussion}). In principle, any star
formation history could be used; an exponentially declining function
is a common choice, for example \citep[see, e.g.,][]{Ruhland09}. We use the
\citet{BC03} stellar population synthesis models with a Chabrier initial mass function (IMF) and the Padova 1994 stellar
libraries to determine how the flux changes with time for each
population of stars that form and take the loss of material back into the interstellar medium into account. We interpolate between the
simple stellar population evolutionary tracks that are provided for six different metallicities to determine
the appropriate evolution given each galaxy's metallicity, calculated using Equation~(\ref{eqn:mz}). 

The star formation rate is normalized so that the total mass in stars remaining at the time of the merger is equal to the stellar mass of the galaxy given by the SAM at that time. The additional burst mass is given by the cold gas mass of the merging galaxy, assuming that all the gas is used up in a burst lasting 1~Gyr. The normalization of the MF was found to be too low compared to observations if no burst was included. The gas mass can be a substantial fraction of the baryonic mass, particularly at high redshifts, and for low and intermediate galaxies even at $z<1$. We test how
different star formation histories and the addition of bursts affect the results in Section~\ref{sec:discussion}. 

Major wet merger remnants redden and fade after the merger, moving onto the red sequence some time after star formation has stopped.  In order to compare directly with the observed distribution, we use the same color cut that was used by \citet{Brown07} to determine which of the model galaxies are on the red sequence at the time of observation, as described in more detail in Section~\ref{sec:results}. With the assumptions described above, the model red sequence approximately matches both the slope and normalization of the observed color--magnitude relation at $z\sim1$ \citep[see Section~\ref{sec:results}]{Brown07,
  Bell04}. Having set up a fairly
realistic population of galaxies at this redshift we explore how
different merging scenarios and star formation histories affect the
evolution of the population between $z=0.9$ and $z=0.1$, where observational results are readily available for comparison. We present the
results for three illustrative models for the evolution of the red
sequence over the last 8 billion years, pictured in
Figure~\ref{fig:cartoon}. We follow the procedure outlined above for all three models until $z\sim1$ and then allow the models to diverge as described below. 

\subsection{Model I}
The first situation we consider, Model I, is how the red sequence evolves if star formation in red sequence galaxies is shut off at $z \sim 1$ and there are no further mergers of any kind. We do not allow any further additions to the red sequence through the shutting down of star formation in blue galaxies. The galaxies already on the red sequence merely evolve passively, becoming fainter and redder, as shown in the first panel of Figure~\ref{fig:cartoon}. The MF does not evolve because the number density of galaxies on the red sequence is not increased through the addition of wet merger remnants, nor is its composition changed by dry mergers between existing red sequence galaxies. This case is extreme in the context of the standard hierarchical picture of galaxy formation, and is certainly unrealistic at intermediate masses, where there must be an influx of blue cloud galaxies after $z=1$ to account for the observed increase in mass on the red sequence \citep{Bell04, Faber07, Brown07}. At the bright end, which we are most interested in here, it shows how already existing massive early-type galaxies would evolve without undergoing mergers. This amounts to the purely passive evolution often advocated for early-type galaxies, albeit a population in which some galaxies continued forming stars until $z=1$. 

To implement this model we truncate each merger tree at $z=1$. We follow the evolution of the major wet merger remnants  that formed before $z=1$, adding the mass and flux contributions from subsequent mergers only if they also take place before $z=1$. The existing stellar populations evolve passively, with no further star formation, until $z=0.1$.

\subsection{Model II}
In the second case (Model II) we assume that early-type galaxies on the red sequence by $z\sim0$ have had their star formation quenched by $z\sim1$, regardless of the mechanism.  We allow all the mergers predicted by the SAM to occur after $z=1$, but as the star formation in these galaxies has already been truncated, all the mergers occurring thereafter are essentially dry. In this model, which is illustrated in the second panel of Figure~\ref{fig:cartoon}, the red sequence continues to grow through mergers after $z=1$ and more massive galaxies are built up through dry mergers, but the red sequence galaxies we see at low redshift all have old populations that formed at least 8 billion years ago. 

In Model II, major wet mergers are no longer the cause of the quenching of star formation after $z=1$. Galaxies involved in mergers that have not already had their star formation quenched through a major wet merger by this time are assumed to have had a constant star formation rate for 4.5~Gyr from $z_f=4$ to $z=0.9$ and a star formation rate of zero thereafter. With no new source of fuel, this is a plausible timescale on which the fuel initially available for star formation could be used up, but the exact time at which we halt star formation is not critical. This particular value was chosen in order to be able to directly compare the changes in the three models from a common starting point at $z=0.9$.

\subsection{Model III}
In the third model scenario, Model III, shown in the third panel of Figure~\ref{fig:cartoon}, we allow all the mergers predicted by the SAM to occur after $z=1$ and assume that star formation continues until a major wet merger occurs, rather than truncating star formation arbitrarily after 4.5 Gyr. This model is closest to the full SAM. Both dry and wet mergers taking place between $z=1$ and $z=0$ affect the evolution of the number density of red sequence galaxies. Some of the galaxies on the red sequence at $z=0$ will have continued to form stars until relatively recently (the time of their last major wet merger). In this way blue progenitors with young stellar populations are added to the red sequence throughout the last 8 billion years. This model illustrates the effects of progenitor bias.

\section{Resultant color and magnitude evolution}
\label{sec:results}

\begin{figure*}[htbp]
	\begin{center}
	\plotone{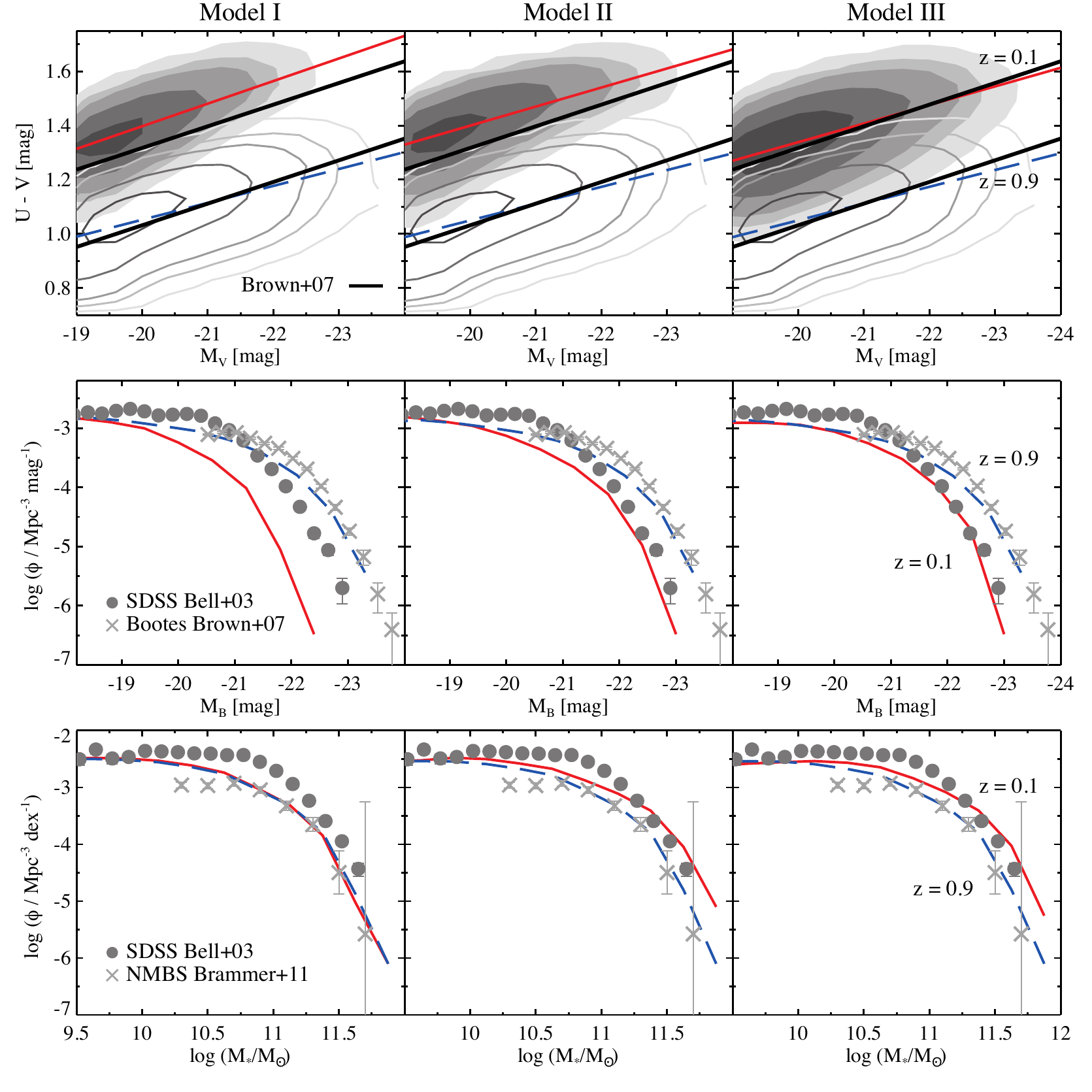}
	\caption[Red sequence evolution for 3 models]{Evolution of the red sequence from $z = 0.9$ (dashed blue lines) to $z = 0.1$ (solid red lines) in the three models. The color--magnitude relation is shown in the top panels, the LF in the middle panels and the MF in the lower panels. The distributions of model galaxies in color--magnitude space are shown by the empty contours ($z=0.9$) and filled contours ($z=0.1$). Linear fits to these distributions are compared to the observed relations from Brown et al. (2007), shown as thick solid lines. At $z=0.1$, the LF and MF are compared to the observed LFs from the SDSS \citep{Bell03}, shown as filled circles. At $z=0.9$, the observed LF is from the Bo\"{o}tes field \citep{Brown07} and the MF from the Newfirm Medium Band Survey \citep[both shown by crosses]{Brammer11}. In the left most panels, galaxies on the red sequence stop forming new stars at $z \sim 0.9$ (4.5 Gyr after formation) and evolve passively, with no further mergers occurring (Model I). In the middle panels, star formation is similarly quenched by $z\sim0.9$ but mergers continue to the present day (Model II). Model III, where star formation may continue after $z=0.9$ until a major wet merger occurs, is shown in the right most panels.  \label{fig:9panel}} 
	\end{center}
\end{figure*}

Figure~\ref{fig:9panel} shows the evolution of the CMR (top panels), LF (middle panels), and MF (lower panels), with each column representing one of the model variations. In all the panels of Figure~\ref{fig:9panel}, the dashed blue lines show the model galaxy distributions at $z=0.9$ and the red solid lines at $z=0.1$. Recall that the models are identical until $z=1$ but differ in their evolution thereafter. For comparison, observational results are shown as gray symbols and thick black lines, as described below.

We compare the model color--magnitude relations at $z=0.9$ and $z=0.1$
with the observed red sequence of galaxies in the Bo\"{o}tes field
\citep{Brown07} parameterized by
\begin{align}
 U - V & = 1.4 - 0.08(M_V-5\log_{10}(h)+20) \nonumber \\
& -0.42(z-0.05)+0.07(z-0.05)^2.
\label{eqn:obscmr}
\end{align}
This is very similar to the CMR found by \citet{Bell04} for galaxies
in the Combo-17 survey. To make a direct comparison, we use the same
magnitude and redshift dependent color cut 0.25 mag below
Equation~(\ref{eqn:obscmr}) to select galaxies on the red sequence in the model. This
is a fairly generous cut that excludes only a small fraction of model
galaxies in the range of $M_V = [-19,-24]$ in the high-redshift bin,
with higher numbers of galaxies excluded toward the faint end. The
distributions of model galaxies in color--magnitude space at $z=0.9$
(open contours) and $z=0.1$ (filled contours) are shown in the top
panels of Figure~\ref{fig:9panel}. The shaded contours represent 1\%, 5\%,
10\%, 25\%, 50\% and 75\% density levels. The observed relations for
galaxies in the Bo\"{o}tes field, given by Equation~(\ref{eqn:obscmr}),
are shown as thick black lines. Although both the observed and model
distributions have some curvature, we compare them by fitting a
straight line to the median of the model galaxy distribution in
magnitude bins of 0.5~mag along the relation at each redshift. The
scatter is calculated with the IDL routine {\it robust\_sigma} using
Tukey's biweight estimator \citep{Beers90}, which is robust to
outliers \citep[see, e.g.,][for measurements of the scatter in the
  observed CMR]{Ruhland09, Bower92, McIntosh05b, Whitaker10}.

We find good agreement between the model and the observed red
sequences at $z=0.9$, but note that both the slope and the
normalization are sensitive to the choice of $M$--$Z$
relation. A shallower $M$--$Z$ relation results in a shallower model
CMR at all redshifts. The scatter in the model red sequence at $z=0.9$ is 0.10~mag, which is somewhat smaller than the
observed scatter (0.21~mag) measured by \citet{Ruhland09} at $z\sim1$
in the GEMS survey. The scatter decreases toward the bright end --- for
massive galaxies ($M_* > 10^{11}\msun$) it is 0.08~mag. This is
also smaller than the measurement from the Newfirm Medium Band Survey (NMBS)
at $z\sim1$ \citep{Whitaker10}. The dispersion in the CMR is largely
driven by the spread in the ages of the stellar populations, as the
model red sequence has been slowly built up by the quenching of star
formation at different times. Allowing for a scatter of 0.1 dex in the
$M$--$Z$, as measured by \citet{Gallazzi06},
increases the scatter at all redshifts, but still does not cause it to
exceed the observed scatter.

In the middle row of Figure~\ref{fig:9panel} we show the $B$-band LFs. Observed LFs of early-type galaxies selected by color from the SDSS \citep{Bell03} at $z \sim 0.1$ and the Bo\"{o}tes field \citep{Brown07} at $0.8<z<1$ are shown with gray circles and crosses, respectively. The color cuts we use to define the red sequence are close to the cuts used by \citet{Brown07}  and \citet{Bell04}, but membership of the model red sequence is largely determined by merger history. As an indication of where the low-redshift LF lies, we show the LF of early-type galaxies selected by color from the SDSS early data release \citep{Bell03}, corrected to $H_0 = 70$~km~s$^{-1}$~Mpc$^{-1}$. The $g$-band of the SDSS has been assumed to be directly comparable to the $B$-band in the Vega system used in this work, and is in good agreement with the LF estimate from the 2dFGRS \citep{Madgwick02}. Note that the observed LFs from the COMBO-17 \citep{Bell04} and DEEP2 \citep{Faber07} surveys have lower normalizations than the \citet{Brown07} LF at $z\sim0.9$, and agree better with the model at the knee of the LF. The bright ends are very similar in all three works.
 
The model and observed MFs at $z=0.9$ and $z=0.1$ are shown in the lower panels of Figure~\ref{fig:9panel}. The MF of $z\sim0.1$ SDSS galaxies selected to be early-type based on a color cut is shown by the filled circles \citep{Bell03}. At high redshift ($0.8<z<1.4$) we show a recent determination of the MF of red galaxies from the NMBS \citep[][]{Brammer11}. This agrees very well with the Schechter function fit to the MF of red sequence galaxies at $0.8 < z < 1$ from \citet{Ilbert10}. The uncertainties in the observational results for the most massive galaxies ($M_* \ga 3 \times 10^{11}\msun$) are large, making it difficult to draw a robust conclusion on their evolution. 

The model and observed number densities at the
bright end agree fairly well at both redshifts, however the
distribution of model galaxies still differs in shape from the observed
distribution, with too few intermediate luminosity galaxies at low redshift. We are
largely focused on the differences in the {\it evolution} of the
bright end of the LF between the three models, to demonstrate the
effects of mergers, rather than trying to match the entire
distribution of galaxies, yet it is important that the space density
of galaxies as a function of luminosity and mass in the models is
approximately correct at our starting point ($z=0.9$). It was necessary to include a burst of star formation at the time of the merger to convert the remaining cold gas mass into stars, in order to match the observed MF. This makes little difference to the evolution of the massive end of the red sequence after $z=1$. We discuss the
impact of bursts of star formation and implications of the differences between the model and observed LFs in
Sections~\ref{sec:sfh} and \ref{sec:LFdisc}.

In Model I, which has only the passive aging of existing stars and no mergers (left panels of Figure~\ref{fig:9panel}), there is rapid evolution of the color and luminosity of galaxies, resulting in a much fainter and redder population by $z=0$. The resulting model CMR (upper left panel) is significantly redder than both the observed relation at $z=0.1$ and the other models. The scatter in the relation measured for $M_V < -19$ is small (0.05~mag). The number density of galaxies as a function of mass does not change after $z=1$ (lower left panel). Despite the constancy of the MF, the LF undergoes significant evolution (middle left panel) due to the rapid fading of the stellar populations, which may have continued to form stars until as late as $z=0.9$ in some galaxies. There are too few bright red galaxies remaining at low redshift, with no increase in mass through merging to counteract the fading of the stellar populations. This gives an upper limit for the amount of luminosity and color evolution; there may be galaxies made up of even older stellar populations that evolve more slowly and galaxies that deplete their gas and move onto the red sequence without undergoing mergers, which do not contribute to the red sequence in these models.
 
In the middle column of Figure~\ref{fig:9panel}, the evolution of the model with mergers but no star formation after $z=0.9$ (Model II) is shown. Although there are still no new stars formed, the fading and reddening of the CMR is less dramatic than in Model I because of the effects of mergers on both the mass and color evolution. While dry mergers increase the mass of galaxies, the slope of the CMR implies that remnants will be no redder than the most massive progenitor, reducing the slope at the bright end of the CMR (see SBS09). The CMR at $z=0.1$ is in better agreement with the observed relation than Model I, although there is still slightly more evolution in color at the bright end than observed. The scatter for galaxies with $M_V < -19$ at $z=0.1$ is 0.06~mag.  In contrast to Model I, the MF of Model II evolves between $z=1$ and $z=0$ due to the redistribution of mass amongst red sequence galaxies and the addition of new major merger remnants to the red sequence. The effect of mergers is to shift the MF to higher masses (lower middle panel). The noticeable growth in the number density of massive galaxies caused by mergers in the models is consistent with the observations. The evolution of the LF incorporates both the passive fading of the stellar populations and the increase in number density of massive galaxies from merging. These act in opposite senses so that the resultant change in the LF (central panel) is much less marked than in Model I. 

In Model III, major merger remnants added to the red sequence after $z=0.9$ are assumed to form stars until the time of the merger if their progenitors have a high gas fraction. Such galaxies have younger luminosity-weighted ages and bluer colors than galaxies with stellar populations that all formed before $z=0.9$. The resulting CMR at $z=0.1$ is in excellent agreement with the observed relation (upper right panel). The scatter in $U-V$ color for galaxies in Model III with $M_V < -19$ at $z=0.1$ is 0.09~mag, which is slightly smaller than the scatter measured for SDSS galaxies at low redshift \citep{Ruhland09}. The increase in scatter compared to Model I is in line with the expectation for a steady influx of blue cloud galaxies onto the red sequence after $z=1$ \citep{vanDokkumFranx01,Ruhland09, Harker06}. For galaxies with $M_* > 10^{11}\msun$ the scatter in $U-V$ color is 0.07~mag, consistent with the scatter measured for galaxies in this mass range in clusters out to $z=0.84$ \citep{vanDokkum08}.

The MF in Model III evolves in the same way as for Model II.  Although some of the galaxies have a different star formation history, the same total mass in stars is added to the red sequence. Galaxies that formed stars after $z=0.9$ may undergo further (dry) mergers once they are on the red sequence. In this way younger stellar populations are incorporated into more massive galaxies. The fading and reddening of the stellar populations is compensated by the addition of mass from mergers, as well as recent additions to the red sequence. The changes at the bright end of the LF are smaller than either of the other models. The resulting evolution of the luminosity function is very mild, although there is substantial growth through merging after $z=1$. This gives an alternative explanation for the mild evolution previously interpreted as the result of purely passive fading of galaxies that formed their stars at high redshifts.

\section{Implications for the evolution of massive galaxies}

\begin{figure}[t]
	\plotone{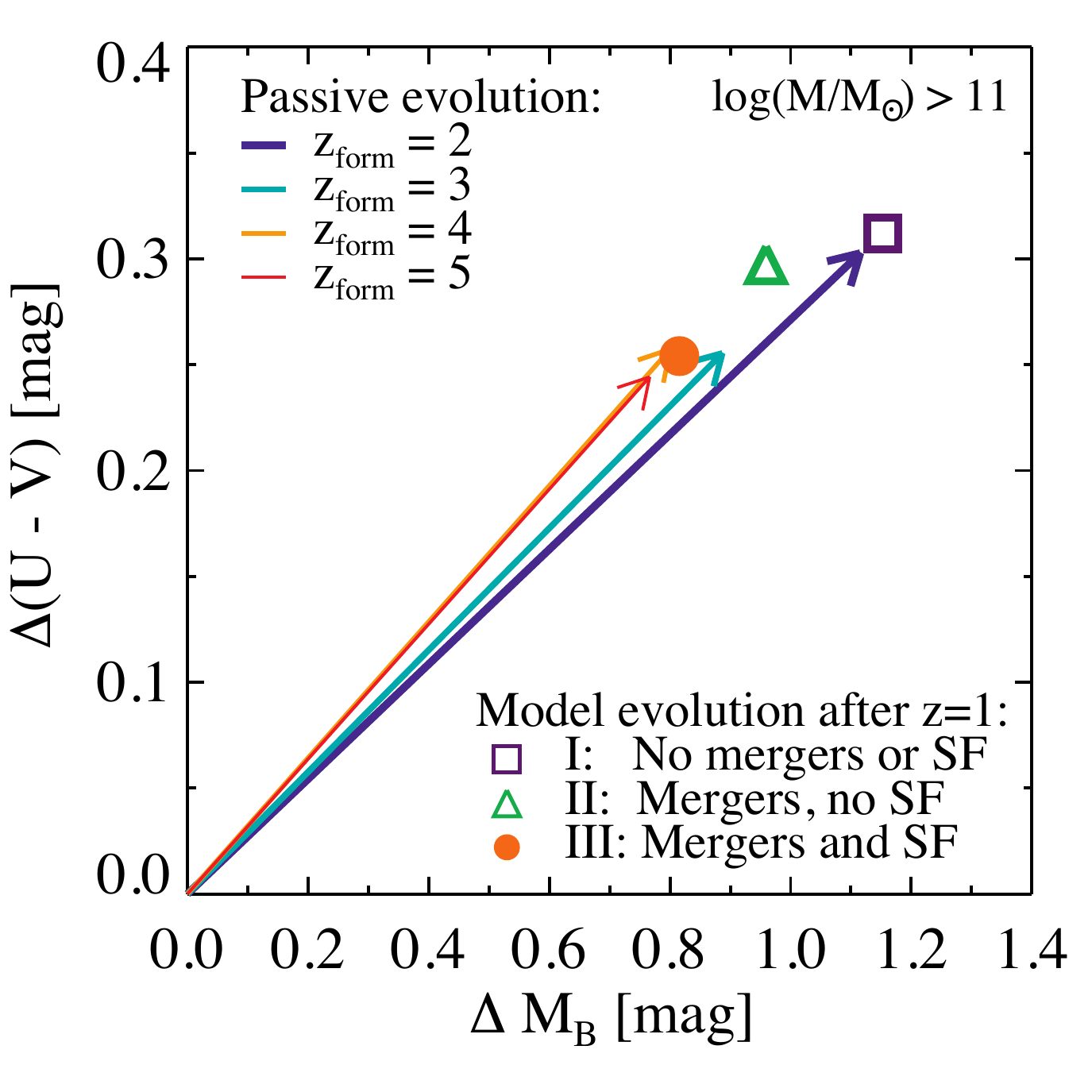}
	\caption[Changes in color and magnitude from $z=0.9$ to $z=0.1$]{Change in color and magnitude from $z = 0.9$ to $z = 0.1$ for galaxies with masses $> 10^{11}{{M_{\sun}}}$ in the  models compared to the expected changes for passively evolving SSPs formed at redshifts from 2 to 5 (arrows). Model I, II, and III are shown by the open square, filled circle and filled triangle, respectively. \label{fig:offsets}} 
\end{figure}

Figure~\ref{fig:offsets} summarizes the changes in color and magnitude between $z=0.9$ and $z=0.1$ for model galaxies with $\log(M/\msun) \gt 11$. We compare the predicted evolution of the three models with the average changes in color and magnitude for passively evolving simple stellar populations (SSPs) in the same mass range, with four choices of formation redshift, $z_f = \left\{2,3,4,5\right\}$. To calculate the expected evolution of the passively evolving population, we use the BC03 stellar population models to simulate 1000 model galaxies drawn from the observed MF for each formation redshift. Rather than simply assuming solar metallicity, we allow for the variation of metallicity with mass, taking into account the relative number of galaxies as a function of mass by assuming they are distributed according to a Schechter function (this is important only at the massive end, where the change in number density with mass is very rapid). We use the observed MF of red galaxies from the SDSS \citep{Bell03}, with parameters $\phi^* = 0.0107 / h^3{\rm{ Mpc}}^{-3} \log_{10} {{M}}^{-1}$, $\log_{10}\left(\frac{M^* h^2}{{M_{\sun}}}\right) = 10.50$ and $\alpha = -0.70$ to assign each galaxy a weight. This has been converted to a Chabrier IMF by subtracting 0.1 dex from the mass \citep{Borch06}. The metallicity of each galaxy is calculated using Equation~(\ref{eqn:mz}), where a mass of $\log(M/\msun) = 10.4$ corresponds approximately to solar metallicity. The amount of evolution found for passively evolving galaxies does not depend strongly on the mass range (i.e., metallicity). The color evolution is approximately that of a solar metallicity SSP, but there is less evolution in the $B$-band, as expected for higher metallicity SSPs. 

In Model I, the population of galaxies in this fixed mass range ($\log(M/\msun) \gt 11$) is the same at both redshifts, whereas in the other models galaxies can move into the sample as their masses are increased through merging. The mass selection thus results in a much smaller sample at $z=0.1$ in Model I than in the other models. The number density of galaxies with $\log(M/\msun) \gt 11$  grows by approximately a factor of two in the models where merging has continued after $z=1$. The population of galaxies is bluer and brighter at $z=0.1$ than the corresponding population of galaxies in Model I. Additions to the red sequence after $z=1$ enhance the effect further (Model III). The relatively small resulting changes in color and magnitude resemble those of a passively evolving population that formed at higher redshift, as can be seen from the similarity in the position of the Model III point and the passive evolution vectors with $z_f$ between 3 and 5. The change in color and magnitude for Model I replicates that of a more recently formed, passively fading population, because there is no merging or additional star formation that can compensate for the rapid luminosity and color evolution. 

Observational measurements of the change in $M/L_B$ can be compared to the change in magnitude at a fixed mass in the models, shown in Figure~\ref{fig:offsets}. For {\it cluster} galaxies with $\log (\mstar/\msun) > 11$  $\Delta M/L_B$ from $z=0.83$ to $z=0$ lies in the range of 0.8 - 1~mag \citep{vanDokkum98, vDvdMarel07,Holden05, Wuyts04, vanDokkumStanford03,vanderWel05}, while $\Delta U-V \sim 0.21$~mag \citep{vanDokkum08}.  Most of the evolution measurements for clusters extrapolate between a single cluster at high $z$ (MS 1054-03 at $z=0.83$) and the Coma cluster at $z=0.022$. The change in $M/L_B$ is larger when the sample is extended to lower masses \citep{Holden10}. For field galaxies, there is greater spread in the measurements of $d\log(M/L_B)/dz$ from the Fundamental Plane \citep[][and references therein]{Treu05,vanderWel05}. For $\Delta z = 0.8$, as plotted in Figure~\ref{fig:offsets}, the change in magnitude measured for field galaxies lies in the range of 1 - 1.9~mag \citep{Rusin05, vanDokkumEllis03, vanderWel04,Gebhardt03, Treu05}. Many of these studies are limited by small sample sizes. The evolution in luminosity and color for all three models fall within the observed range, but more work is needed to pin down the observed evolution for a larger sample of field galaxies. 

A fairly straight-forward way of measuring the evolution of the LF (provided that a Schechter function or other analytic form fits the LF reasonably well) is by determining the change in magnitude at a fixed space density \citep[e.g.,][]{Brown07}. This has the advantage that it is relatively insensitive to the details of red galaxy selection and avoids the use of a magnitude threshold that is sensitive to the exact shape of the LF at the bright end. 

\begin{figure}[t!]
\centering
\plotone{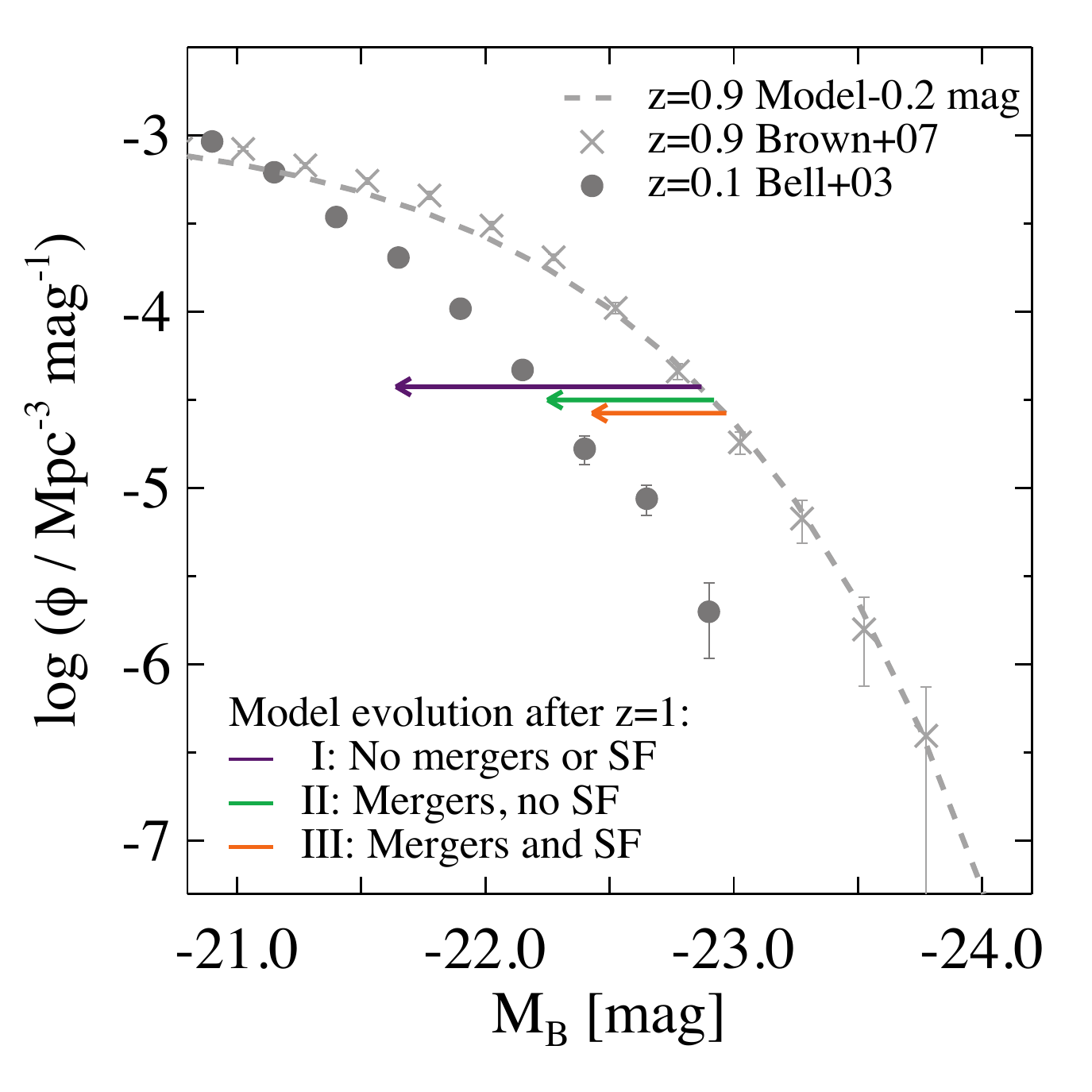}
\caption[Changes in magnitude at fixed space density]{Change in magnitude from $z = 0.9$ to $z = 0.1$ at a fixed space density of $10^{-4.5} {\rm Mpc^{-3} mag^{-1}}$ in the models, compared to the observed evolution of the LF. The low-$z$ LF data (filled circles) are from the SDSS \citep{Bell03}, while the high-$z$ LF data points (crosses) are from the Bo\"{o}tes field \citep{Brown07}. A Schechter function fit to the model LF at $z=0.9$, shifted by 0.2~mag to coincide with the Bo\"{o}tes data at a space density of $\phi = 10^{-4.5} {\rm Mpc^{-3} mag^{-1}}$, is shown by the thick dashed line. The three models have been slightly offset in space density for clarity. The luminosity evolution of Model I (blue arrow) far exceeds the observed evolution. Models II (green arrow) and III (orange arrow) agree well with observations. \label{fig:lfevol}} 
\end{figure}

In Figure~\ref{fig:lfevol} we show the change in magnitude for the three
models at a space density of $10^{-4.5} {\rm Mpc^{-3} mag^{-1}}$,
overlaid on the observed LFs from the Bo\"{o}tes field \citep[NOAO
  Deep Wide-Field Survey;][]{Brown07} at $z=0.9$ and the SDSS
\citep{Bell03} at $z=0.1$. The evolution vectors for Model I and III
have been offset slightly for clarity. The characteristic magnitudes
of the model Schechter functions at both redshifts were shifted by
0.2~mag so that the model fit at $z=0.9$ (thick dashed line) intersects with the
Bo\"{o}tes data at the space density of
interest. 

Figure~\ref{fig:lfevol} emphasizes that a red sequence galaxy population with a realistic mix of stellar populations at $z=1$ will undergo very rapid evolution in magnitude at fixed space density if there is no growth in mass (Model I, blue arrow). The evolution is dramatically slowed by growing the red sequence population through dry mergers between already-existing early-type galaxies to larger masses (Model II, green arrow), and slowed further by the addition of wet merger remnants that have recently had their star formation quenched (Model III, orange arrow). The luminosity evolution in models that include mergers is reasonably consistent with the observations at all space densities probing the bright end of the LF.  

\section{Discussion}
\label{sec:discussion}

The behavior reported in this paper is remarkably robust to a number of the model choices we have made. Though some of the galaxy properties are taken directly from the SAM, and thus depend on the parameters and prescriptions used there, others are determined independently in this work. Choices made in the three models presented here include the $M$--$Z$ relation, the mass ratio used to separate major and minor mergers, the gas fraction threshold below which a merger is considered gas-rich, the gas fraction threshold above which a galaxy remains in the blue cloud, the time at which star formation starts, the normalization and shape of the star formation history, the stellar population models and associated stellar library, and the IMF. In Model I there is the additional assumption of a time after which no mergers take place ($z=1$) and for Model II, a time after which any ongoing star formation stops (4.5 Gyr).

\subsection{Merger histories}
We implement our model using one particular incarnation of a hierarchical merger tree, taken from the S08 SAM, but expect the results to hold more generally in standard hierarchical models. The merger statistics of various models, including S08, were compared in \citet{Jogee09} and \citet{Lotz11}. There is broad qualitative agreement between the models, but in detail the merger properties can differ substantially. This is most likely due to the different implementations of baryonic physics, rather than differences in the merger statistics of the dark matter halos, as no significant differences in the results of the SAM were found when the model was implemented within $N$-body merger trees instead.  

In our simple model, the gas fractions and mass ratios of merging galaxies determine which galaxies move onto the red sequence.  The number density of red galaxies then depends critically on the choice of model and the parameters used to select which galaxies are quenched, but we expect the relative differences between the three scenarios to hold using other merger trees. It is reassuring to note that the baryonic mass ratio distributions of merging galaxies in the S08, \citet{Croton06} and \citet{Stewart09b} models are similar \citep{Lotz11}. The mean gas fraction of merging galaxies in the S08 SAM is fairly high but comparable to \citet{Stewart09b}. The importance of mergers of different gas fractions differs from model to model as a function of time. In the S08 model major mergers at $z<1$ are mostly gas-poor, while mergers with a higher mass ratio usually involve a more gas-rich lower mass galaxy and so minor and intermediate mass ratio mergers are gas-rich until lower redshift. This is not necessarily the case in other models (see Figure~7 in \citealt{Lotz11}). 

The merger fraction for the S08 model was found to compare well with the observed fraction of morphologically disturbed galaxies in \citet{Jogee09} and with the major close pair fraction in \citet{Lotz11} (see also \citealt{Bell06b,Bell06a} for comparisons with earlier versions of the model).

\subsection{Mass--Metallicity relation}

The normalization and slope of the color-magnitude relation produced by the model are strongly influenced by the $M$--$Z$ relation, but also depend on the star formation histories of the galaxies. A linear fit to  the observed relation from \citet{Gallazzi06} results in good agreement with the slope of the observed red sequence and a small offset in color compared to observations. We adjusted the fitted $M$--$Z$relation by 0.004 (see Equation~(\ref{eqn:mz})) to improve the match at $z=0.9$.  The mean color of the red sequence is slightly redder and no offset to the $M$--$Z$-fit is required if we do not incorporate significant bursts of star formation but assume only a constant star formation history for each galaxy (see Section~\ref{sec:sfh}). The consequences of dry mergers for the colors of massive red galaxies depend on the existence of a tilt in the color-magnitude relation but are fairly insensitive to the exact choice of $M$--$Z$ relation.

The scatter in the red sequence is driven mainly by age differences. The scatter in all three models is smaller than observed. This leaves room to include scatter in the $M$--$Z$ relation. Assuming a log normal relation with a width of 0.1~dex \citep{Tremonti04,Gallazzi06} results in slightly larger scatter in all the model red sequences, still well within the range expected from observations. The differences between the models are evident even when improbably large scatter in the $M$--$Z$ relation is used. 

One could imagine that if all mergers back to the highest redshifts are considered, most dark matter halos are likely to have had a major merger, with many of these mergers occurring at early times when the masses of the galaxies involved were very small. Using a non-evolving $M$--$Z$ relation then implies that the metallicities would be very low. With no further star formation, the metallicities of remnant galaxies on the red sequence would remain unrealistically low. This turns out not to be a concern. We consider mergers taking place after the assumed formation redshift of $z_f = 4$. Furthermore, we assume that only the final gas-rich major merger on any branch of the merger tree moves the remnant onto the red sequence. This excludes some very gas rich mergers occurring at high redshifts, which is a reasonable assumption as long as the remnant retains sufficient gas to remain in the blue cloud or re-accretes gas before a second major merger. It does lead to a concern that similar types of mergers may have different effects depending on when they occur, or be entirely ignored in one case while creating a red sequence galaxy in another case. An alternative approach could be to enforce an upper gas fraction threshold, above which the remnant is assumed to remain in the blue cloud (see Section \ref{sec:LFdisc} below). These conceptual issues largely affect the creation of the model red sequence but will have little effect on the differences between the models at low redshift. We have shown that the red sequence produced by $z=1$ under the current set of assumptions agrees sufficiently well with observations to merit its use as a reference.

\subsection{SF histories}
\label{sec:sfh}
We have chosen a very simple constant star formation history for all galaxies as the default. This was also the approach followed by \citet{Kriek08} in their models of red sequence color evolution. There are many other options that could be considered. The scatter we find is consistent with that of \citet{Ruhland09}, who used an exponentially declining SFR in their exploration of how the scatter in the CMR is affected by the continuous addition of quenched blue cloud galaxies. An exponentially declining SFH may not be a good description for all galaxies, however. Lyman break galaxies that are forming stars at high levels at redshifts above two may be better described by an increasing SFH, for example \citep{Lee10,Papovich11}. Previous work on the progenitor bias \citep[]{vanDokkumFranx01} used three parameters to describe the star formation history --- the start time, end time, and a parameter that describes how SF varies within this period. They note that the  details of how the SFR changes with time are not very important; the evolution of an individual galaxy can be approximately described as the evolution of an SSP that formed at the time given by the mean luminosity-weighted age of the galaxy \citep[][]{vanDokkum98b,vanDokkumFranx01}. 

In our fiducial model, star formation for all galaxies begins at a redshift of four. We tested more extreme starting times corresponding to $z=2$ and $z=8$, as well as a variation in which the starting times are randomly chosen from a Gaussian distribution of width 0.5~Gyr centered on $z=4$ and constrained to lie between $z=10$ and $z=1.5$. For a more recent starting time ($z_f = 2$) the model red sequence is bluer than the observed relation by approximately 0.05~mag at $z=0.9$. At $z=0.1$, the red sequence for Model I is slightly redder than observed, Model II matches well with the observed relation, and Model III remains slightly bluer than observed. The luminosity evolution for all three models is faster as all the stars have formed more recently. This gives larger changes in both color and magnitude between $z=0.9$ and $z=0.1$. The relative changes for the three models are in the same sense as before but the impact of allowing star formation to continue after $z=1$ is lessened, leading to a smaller difference between Model II and III. When star formation starts at $z=8$, the evolution is slightly slower for all models but the results are very similar to the fiducial case. The results when a range of starting times is allowed are almost indistinguishable from those with a single starting time.

If major wet mergers are the catalyst for the quenching of star formation, as we have assumed in our models, they are likely to be accompanied by a burst of star formation that uses up or expels the remaining fuel from the central regions of the galaxy. We found that it was necessary to convert the significant amounts of cold gas in major wet merger progenitor galaxies in the SAM to stars in order to produce high enough masses. It is not surprising that neglecting the cold gas results in lower normalization, as this gas is converted into stars in the full SAM, and the MF of the SAM at $z=0$ matches well with observations. Bursts of star formation have stronger impact at higher redshifts, where the galaxies have higher gas fractions, approximately doubling the mass of galaxies with $M_* \la 3 \times 10^{10}\msun$. We tested other ways of implementing bursts as well as star formation histories in which the same amount of mass is formed with no burst. It has been shown that star formation in mergers is only mildly enhanced with respect to normal star-forming galaxies at $z\la 1$ \citep[][]{Barton07, Robaina09a, Jogee09}; thus we would not expect a large fraction of the mass to form in merger-induced bursts of star formation. The enhancement in star formation due to mergers in the S08 SAM was found to agree well with observations \citep{Robaina09a}, suggesting that dramatic bursts of star formation are not needed, at least at $z<1$. At higher redshifts, more intense bursts of star formation may occur.  In general, the enhancement in SFR caused by a merger is dependent on the morphologies, masses and gas fractions of the galaxies involved, as well as the orbital parameters and the mass ratio. The relative evolution of the three models and our conclusions remain the same whether the gas mass is excluded, included in a single burst at the time of the merger or incorporated into an extended period of star formation, although the colors do depend on this choice. With a constant star formation rate and no burst, the colors are redder by $\sim 0.1$~mag on average. The color is also dependent on the length of the burst; a shorter, more intense burst of star formation produces slightly bluer galaxies, but again has little impact on our conclusions. Allowing star formation to continue for some time after the merger may be more realistic than the sudden quenching we assume, but it would make little difference to the results.  

It is possible to produce a CMR with a similar slope and normalization at $z=1$ (using the same $M$--$Z$ relation) by assuming that all the stars formed in a single burst at fairly high redshift ($z \ga 2$) rather than over an extended period of time. The scatter in the relation is caused by differences in the color evolution for different metallicity populations rather than differences in age. We find that the scatter at the massive end of the relation in this model is unrealistically small at $z=0.9$ (0.02~mag for $M_*>10^{11}\msun$). Allowing for scatter in the $M$--$Z$-relation would compensate for this to some extent. Although the scatter in the relation as a whole remains approximately constant to low redshifts, the scatter for the most massive galaxies actually increases slightly in this model (0.04~mag at $z=0.1$). The evolution in color and luminosity to low redshifts is too rapid compared to observations, even if dry mergers are included. An earlier formation redshift ($z_f = 4$) results in too red a relation by $z=1$ although the change in color between $z=1$ and $z=0$ is approximately correct. This incompatibility between high redshift single-burst passive evolution models and the observed color evolution has been pointed out by \citet{Kriek08, Whitaker10} and others.

\subsection{The shape of the mass and luminosity functions}\label{sec:LFdisc}

The shape of the model LF and MF differs somewhat from the observed relations (see Figure~\ref{fig:9panel}). The number densities of galaxies at the bright end and faint end in the models that include mergers after redshift one (Model II and III) agree reasonably with the observed number densities from \citet{Brown07} but there is a shortage of intermediate luminosity galaxies. In the models, the evolution in the MFs results solely from mergers, but includes new additions to the red sequence through major mergers, as well as the increase of mass in existing red sequence galaxies. There is very little evolution below the knee of the MF, even in the two models where mergers occur, suggesting that more migration of blue cloud galaxies onto the red sequence than predicted by this model is necessary to match the observed evolution at intermediate masses. Secular processes such as disk instabilities may lead to bulge formation through the development of bars and clumps \citep[e.g.,][]{Efstathiou82, Mo98, Combes00, Cole00, Bower06}. At lower masses, model galaxies appear to be in place too early. 

The masses of merging galaxies are extracted from the S08 SAM, so similar discrepancies in the shape of the MF are also expected for the full SAM. The total MF of the SAM, as well as the MFs split by color, agree well with observations from the SDSS \citep{Bell03} at low redshift. There is very little evolution in the SAM MF, however, resulting in increasing differences to the observed relations at higher redshifts. This seems to be the case in a number of SAMs and is discussed in detail in \citet{Fontanot09}. Although the SAM produces a bimodal distribution of galaxies in color--magnitude space at $z=0$, the distribution also becomes increasingly different from the observed distribution at higher redshifts. Even if the total LF of the SAM matched well with observations at $z\sim1$, the relative contributions from red and blue galaxies would not necessarily agree.  The difficulties in matching the observed LFs in detail were to some extent our motivation for simplifying the model and using only the fairly robust merger trees from the SAM as the basis of this work. Future work using the halo occupation distribution (HOD) to ensure that the distribution of galaxy masses in the model matches that of observations \citep[e.g.][]{Hopkins08a, Conroy07} will allow us to compare quantitative predictions with the observed evolution. In this paper we concentrate on the relative changes between different models at a fixed mass and space density to demonstrate the effects of mergers, rather than making a direct comparison with the observed evolution. 

It has been pointed out that bulges are less likely to form in extremely gas-rich mergers \citep{Hopkins09a, Hopkins09b}.  As less massive galaxies have higher gas fractions, this would inhibit the formation of low-mass red sequence galaxies. This effect is redshift dependent because galaxies had higher gas fractions in the past. The efficiency of bulge formation depends on mass ratio, gas fraction and the orbital parameters of the merger. It is beyond the scope of this work to include this level of detail, however we found that applying a simple upper gas fraction threshold for spheroid formation strongly suppresses the number of low mass red sequence galaxies formed. We find excellent agreement between the observed and model MFs at the low-mass end at $z=0.1$ using an upper gas fraction threshold of 80\%, above which the merger remnants are assumed to stay in the blue cloud rather than moving onto the red sequence. The MF at $z=0.9$ overshoots the observed distribution at low masses, however this is where incompleteness in the observations may be an issue. If no upper threshold is applied, the LF and MF overshoot the observations at low masses. With a threshold of 70\%, the mass functions agree well at $z=0.9$ but more fading of low mass blue cloud galaxies without mergers is required after $z=1$ to reproduce the low-redshift mass function. Adjusting the threshold has almost no impact on the growth at the massive end. 

\subsection{Choice of stellar population model and IMF}

We have verified that the implications of mergers on the evolution of color and luminosity described above are independent of the exact choice of stellar population synthesis model. Using the models of \citet{Maraston05} with a Kroupa IMF results in somewhat slower evolution between $z=1$ and the present day for simple stellar populations, and correspondingly slower evolution in both color and magnitude in all three models. The color--magnitude relation, LF and MF are similar to those shown in Figure~\ref{fig:9panel}, and the differences in evolution between the models at low redshifts are in the same sense as shown in Figure~\ref{fig:offsets}. The main difference between the \citet{BC03} and \citet{Maraston05} models is the treatment of post-main sequence evolution, particularly TP AGB stars. The impact is strongest in the near infrared bands and at high redshifts, as shown by \citet{vanderWel06} and others, and so we do not expect very big differences.

The choice of IMF affects the evolution of both luminosity and color. The distribution of stellar masses has a stronger impact on the luminosity evolution than the color evolution, which is determined largely by the age of the stellar population, however \citep{Tinsley80, vanDokkum08}. The magnitude of the evolution in our models would depend on the IMF, but we expect the differences between the models to hold for any of the ``standard'' IMFs. 

\subsection{Relation to previous works}

The progenitor bias \citep{vanDokkumFranx96, vanDokkum00,
  vanDokkumFranx01} and consequences of mergers that we have
demonstrated here have been recognized as an important factor for the
evolution of the red sequence for some time and pointed out in
numerous papers on the evolution of massive early-type galaxies
\citep[for e.g.,][]{Brown07,Bell07, Kriek08,
  Whitaker10}. What is new in this work is the calculation of the
magnitude of those effects in a hierarchical model, and a
comparison of the evolution of massive red galaxies with and
without mergers after $z\sim1$.

Models by \citet{Hopkins06, Hopkins06b} considered the co-evolution of early-type galaxies and quasars, assuming a quasar phase is triggered by merging activity and spheroidal galaxies form as remnants of mergers.  They use the observed quasar luminosity function as a starting point and reproduce a number of properties of early-type galaxies, including the LF, MF and evolution of the $M/L$ ratio. Later models using the HOD approach with the simple assumption that major mergers are responsible for quenching star formation and forming spheroids also predicted a number of properties of early-type galaxies that match well with observations \citep{Hopkins08a,Hopkins08b}. We have used a simpler model framework but make similar assumptions on the formation of spheroids. Our results are in good agreement. 

The combined effects of mass build up through mergers and stellar evolution that we test here are naturally included in more complex SAMs \citep[see, e.g.,][]{Somerville08, Bower06, DeLucia07, Monaco07}. Many of these models have implicitly or explicitly shown that massive early-type galaxies form hierarchically rather than evolving purely passively for many years now. In particular, it has been shown that the fairly rare massive early-types in the densest environments (brightest cluster galaxies) continue assembling to low redshifts, mostly through dry mergers, although the stars in these merging components may have formed at high redshifts \citep{Aragon-Salamanca98, DeLucia07}. In a recent paper \citet{DeLucia11} examine how much mass in bulge galaxies has been contributed by mergers and disk instabilities. Interestingly, they find that disk instabilities play the most important role at intermediate masses, where the discrepancy between the model and observed MFs is greatest in our merger-only bulge-formation scenario. 

The LF is often used to probe the
mass growth of massive early-types by testing whether there is evidence for any evolution beyond that of an old, passively evolving model \citep[e.g.][]{Wake06, Cimatti06, Cool08}. We have shown
that even with mergers the LF can appear to evolve only by the amount
expected from passive evolution. One of the strongest implications of
this work is then that the LF alone cannot be used to
distinguish between the passive and hierarchical formation scenarios
for massive early-type galaxies. Mass measurements that are
independent of the assumption of passive evolution are
required. A recent measurement of the evolution of the MF
where the LF is converted to a MF using the observed evolution
of the $M/L$ led to a higher estimate of the change in number
density of massive galaxies that is more consistent with the
predictions of hierarchical models \citep{Robaina10}. \citet{Matsuoka10} also measure a greater change in number density using masses determined from spectral energy distribution (SED) fitting, but other works that use similar methods still find very little evolution \citep{Ferreras09, Banerji10}. Note that there is strong model dependence in fitting SEDs to obtain masses, which can lead to large uncertainties \citep[see][]{Gallazzi09b}. 

\section{Conclusions}
\label{sec:summary}

We have used galaxy merger histories from a hierarchical model, in
conjunction with simple star formation histories and stellar
population synthesis models, to explore the color and magnitude
evolution of red sequence galaxies in several scenarios.

We use the galaxy merger trees from the
S08 SAM combined with stellar evolutionary tracks
from the \citet{BC03} model, making simple assumptions on the relation
of mass to metallicity and the redshift at which galaxy formation
begins. We assume that major wet mergers are effective at quenching
star formation but that star formation may also shut down after some
period of time, independently of mergers. With this model setup, we
examine the role of recent mergers and the effect of adding younger
populations to the red sequence at late times.

A red sequence population with a realistic color--magnitude distribution at $z=1$ can be formed through the passive fading of galaxies that form in a single burst relatively recently ($z_f\sim2$) or through more extended star formation, shut down by major mergers at different times, in galaxies that began forming earlier ($z_f \sim 4$). Galaxies that form all their stars at high redshifts have colors that are too red by $z=1$ and produce a CMR that has smaller scatter than observed, but the slow reddening they experience thereafter gives approximately the observed amplitude of color evolution.

A population with approximately the correct color at $z=1$ evolves very rapidly over the last half of cosmic history if no further mergers occur and there is no new mass added to the red sequence (Model I). This results in a color--magnitude relation that is too red at low redshift, with too few bright red galaxies and too much luminosity evolution at fixed space density compared to observations. At a fixed mass, the change in color and magnitude predicted by such a model are approximately the same as the changes predicted for a purely passively evolving population that formed at $z_f\sim2$. 

Dry mergers occurring between $z=1$ and $z=0$ slow down the luminosity and color evolution of the early-type galaxy population. The mass (and light) added through mergers counteracts the fading of the luminosities expected from stellar evolution when no new stars are forming.
The result is a slightly bluer CMR at $z=0$ that is closer to the observed
relation, and more realistic evolution of the LF and MF (Model II).
By allowing star formation to continue in some galaxies after $z=1$,
the changes in a population's color and luminosity are reduced even
further (Model III). This late morphological transformation is known
as progenitor bias \citep{vanDokkumFranx01}. Both the CMR and the LF
evolution of this model are consistent with the observed
evolution. The smaller changes in color and magnitude at a fixed mass
replicate the behavior of a passively evolving population that formed
in a single burst at high redshift ($z_f = 3 - 5$). 

At the bright end of the red sequence, the seemingly slow passive evolution of luminosity and color is equally well interpreted as the passive evolution of an ancient population, or as the cosmologically motivated hierarchical growth of an evolving population.  Although they predict very similar evolution, the implications of these scenarios for the growth of stellar mass are very different. Diagnostics other than the LF, such as the number density evolution estimated using dynamical $M/L$ information, the growth rate determined from close-pair statistics and the evolution of the clustering strength of early-type galaxies lend support to the hierarchical picture. 
 
\section*{}
We thank Hans-Walter Rix, Arjen van der Wel, and David Wake for useful conversations and comments, and Gabriel Brammer for making his data available to us in electronic format. We thank the referee for a constructive report that has greatly improved the paper. This research has made extensive use of the IDL Astronomy Library (http://idlastro.gsfc.nasa.gov/) and NASA's Astrophysics Data System Bibliographic Services.

\bibliographystyle{apj}

\begin{thebibliography}{128}
\expandafter\ifx\csname natexlab\endcsname\relax\def\natexlab#1{#1}\fi

\bibitem[{{Andreon}(2011)}]{Andreon11}
{Andreon}, S. 2011, \aap, 529, L5

\bibitem[{{Aragon-Salamanca} {et~al.}(1998){Aragon-Salamanca}, {Baugh}, \&
  {Kauffmann}}]{Aragon-Salamanca98}
{Aragon-Salamanca}, A., {Baugh}, C.~M., \& {Kauffmann}, G. 1998, \mnras, 297,
  427

\bibitem[{{Arnouts} {et~al.}(2007){Arnouts}, {Walcher}, {Le F{\`e}vre},
  {Zamorani}, {Ilbert}, {Le Brun}, {Pozzetti}, {Bardelli}, {Tresse}, {Zucca},
  {Charlot}, {Lamareille}, {McCracken}, {Bolzonella}, {Iovino}, {Lonsdale},
  {Polletta}, {Surace}, {Bottini}, {Garilli}, {Maccagni}, {Picat},
  {Scaramella}, {Scodeggio}, {Vettolani}, {Zanichelli}, {Adami}, {Cappi},
  {Ciliegi}, {Contini}, {de la Torre}, {Foucaud}, {Franzetti}, {Gavignaud},
  {Guzzo}, {Marano}, {Marinoni}, {Mazure}, {Meneux}, {Merighi}, {Paltani},
  {Pell{\`o}}, {Pollo}, {Radovich}, {Temporin}, \& {Vergani}}]{Arnouts07}
{Arnouts}, S., {et~al.} 2007, \aap, 476, 137

\bibitem[{{Baldry} {et~al.}(2004){Baldry}, {Glazebrook}, {Brinkmann},
  {Ivezi{\'c}}, {Lupton}, {Nichol}, \& {Szalay}}]{Baldry04}
{Baldry}, I.~K., {Glazebrook}, K., {Brinkmann}, J., {Ivezi{\'c}}, {\v Z}.,
  {Lupton}, R.~H., {Nichol}, R.~C., \& {Szalay}, A.~S. 2004, \apj, 600, 681

\bibitem[{{Banerji} {et~al.}(2010){Banerji}, {Ferreras}, {Abdalla}, {Hewett},
  \& {Lahav}}]{Banerji10}
{Banerji}, M., {Ferreras}, I., {Abdalla}, F.~B., {Hewett}, P., \& {Lahav}, O.
  2010, \mnras, 402, 2264

\bibitem[{{Barnes} \& {Hernquist}(1996)}]{Barnes96}
{Barnes}, J.~E., \& {Hernquist}, L. 1996, \apj, 471, 115

\bibitem[{{Barton} {et~al.}(2007){Barton}, {Arnold}, {Zentner}, {Bullock}, \&
  {Wechsler}}]{Barton07}
{Barton}, E.~J., {Arnold}, J.~A., {Zentner}, A.~R., {Bullock}, J.~S., \&
  {Wechsler}, R.~H. 2007, \apj, 671, 1538

\bibitem[{{Beers} {et~al.}(1990){Beers}, {Flynn}, \& {Gebhardt}}]{Beers90}
{Beers}, T.~C., {Flynn}, K., \& {Gebhardt}, K. 1990, \aj, 100, 32

\bibitem[{{Bell} {et~al.}(2003){Bell}, {McIntosh}, {Katz}, \&
  {Weinberg}}]{Bell03}
{Bell}, E.~F., {McIntosh}, D.~H., {Katz}, N., \& {Weinberg}, M.~D. 2003, \apjs,
  149, 289

\bibitem[{{Bell} {et~al.}(2006{\natexlab{a}}){Bell}, {Phleps}, {Somerville},
  {Wolf}, {Borch}, \& {Meisenheimer}}]{Bell06b}
{Bell}, E.~F., {Phleps}, S., {Somerville}, R.~S., {Wolf}, C., {Borch}, A., \&
  {Meisenheimer}, K. 2006{\natexlab{a}}, \apj, 652, 270

\bibitem[{{Bell} {et~al.}(2007){Bell}, {Zheng}, {Papovich}, {Borch}, {Wolf}, \&
  {Meisenheimer}}]{Bell07}
{Bell}, E.~F., {Zheng}, X.~Z., {Papovich}, C., {Borch}, A., {Wolf}, C., \&
  {Meisenheimer}, K. 2007, \apj, 663, 834

\bibitem[{{Bell} {et~al.}(2004){Bell}, {Wolf}, {Meisenheimer}, {Rix}, {Borch},
  {Dye}, {Kleinheinrich}, {Wisotzki}, \& {McIntosh}}]{Bell04}
{Bell}, E.~F., {et~al.} 2004, \apj, 608, 752

\bibitem[{{Bell} {et~al.}(2006{\natexlab{b}}){Bell}, {Naab}, {McIntosh},
  {Somerville}, {Caldwell}, {Barden}, {Wolf}, {Rix}, {Beckwith}, {Borch},
  {H{\"a}ussler}, {Heymans}, {Jahnke}, {Jogee}, {Koposov}, {Meisenheimer},
  {Peng}, {Sanchez}, \& {Wisotzki}}]{Bell06a}
---. 2006{\natexlab{b}}, \apj, 640, 241

\bibitem[{{Benson} {et~al.}(2003){Benson}, {Bower}, {Frenk}, {Lacey}, {Baugh},
  \& {Cole}}]{Benson03}
{Benson}, A.~J., {Bower}, R.~G., {Frenk}, C.~S., {Lacey}, C.~G., {Baugh},
  C.~M., \& {Cole}, S. 2003, \apj, 599, 38

\bibitem[{{Bernardi} {et~al.}(2007){Bernardi}, {Hyde}, {Sheth}, {Miller}, \&
  {Nichol}}]{Bernardi07a}
{Bernardi}, M., {Hyde}, J.~B., {Sheth}, R.~K., {Miller}, C.~J., \& {Nichol},
  R.~C. 2007, \aj, 133, 1741

\bibitem[{{Bezanson} {et~al.}(2009){Bezanson}, {van Dokkum}, {Tal},
  {Marchesini}, {Kriek}, {Franx}, \& {Coppi}}]{Bezanson09}
{Bezanson}, R., {van Dokkum}, P.~G., {Tal}, T., {Marchesini}, D., {Kriek}, M.,
  {Franx}, M., \& {Coppi}, P. 2009, \apj, 697, 1290

\bibitem[{{Blanton}(2006)}]{Blanton06}
{Blanton}, M.~R. 2006, \apj, 648, 268

\bibitem[{{Blanton} {et~al.}(2003){Blanton}, {Hogg}, {Bahcall}, {Baldry},
  {Brinkmann}, {Csabai}, {Eisenstein}, {Fukugita}, {Gunn}, {Ivezi{\'c}},
  {Lamb}, {Lupton}, {Loveday}, {Munn}, {Nichol}, {Okamura}, {Schlegel},
  {Shimasaku}, {Strauss}, {Vogeley}, \& {Weinberg}}]{Blanton03}
{Blanton}, M.~R., {et~al.} 2003, \apj, 594, 186

\bibitem[{{Borch} {et~al.}(2006){Borch}, {Meisenheimer}, {Bell}, {Rix}, {Wolf},
  {Dye}, {Kleinheinrich}, {Kovacs}, \& {Wisotzki}}]{Borch06}
{Borch}, A., {et~al.} 2006, \aap, 453, 869

\bibitem[{{Bower} {et~al.}(2006){Bower}, {Benson}, {Malbon}, {Helly}, {Frenk},
  {Baugh}, {Cole}, \& {Lacey}}]{Bower06}
{Bower}, R.~G., {Benson}, A.~J., {Malbon}, R., {Helly}, J.~C., {Frenk}, C.~S.,
  {Baugh}, C.~M., {Cole}, S., \& {Lacey}, C.~G. 2006, \mnras, 370, 645

\bibitem[{{Bower} {et~al.}(1998){Bower}, {Kodama}, \& {Terlevich}}]{BKT98}
{Bower}, R.~G., {Kodama}, T., \& {Terlevich}, A. 1998, \mnras, 299, 1193

\bibitem[{{Bower} {et~al.}(1992){Bower}, {Lucey}, \& {Ellis}}]{Bower92}
{Bower}, R.~G., {Lucey}, J.~R., \& {Ellis}, R.~S. 1992, \mnras, 254, 601

\bibitem[{{Boylan-Kolchin} {et~al.}(2008){Boylan-Kolchin}, {Ma}, \&
  {Quataert}}]{Boylan-Kolchin08}
{Boylan-Kolchin}, M., {Ma}, C.-P., \& {Quataert}, E. 2008, \mnras, 383, 93

\bibitem[{{Brammer} {et~al.}(2009){Brammer}, {Whitaker}, {van Dokkum},
  {Marchesini}, {Labb{\'e}}, {Franx}, {Kriek}, {Quadri}, {Illingworth}, {Lee},
  {Muzzin}, \& {Rudnick}}]{Brammer09}
{Brammer}, G.~B., {et~al.} 2009, \apjl, 706, L173

\bibitem[{{Brammer} {et~al.}(2011){Brammer}, {Whitaker}, {van Dokkum},
  {Marchesini}, {Franx}, {Kriek}, {Labb{\'e}}, {Lee}, {Muzzin}, {Quadri},
  {Rudnick}, \& {Williams}}]{Brammer11}
---. 2011, \apj, 739, 24

\bibitem[{{Brown} {et~al.}(2007){Brown}, {Dey}, {Jannuzi}, {Brand}, {Benson},
  {Brodwin}, {Croton}, \& {Eisenhardt}}]{Brown07}
{Brown}, M.~J.~I., {Dey}, A., {Jannuzi}, B.~T., {Brand}, K., {Benson}, A.~J.,
  {Brodwin}, M., {Croton}, D.~J., \& {Eisenhardt}, P.~R. 2007, \apj, 654, 858

\bibitem[{{Bruzual} \& {Charlot}(2003)}]{BC03}
{Bruzual}, G., \& {Charlot}, S. 2003, \mnras, 344, 1000

\bibitem[{{Cimatti} {et~al.}(2006){Cimatti}, {Daddi}, \& {Renzini}}]{Cimatti06}
{Cimatti}, A., {Daddi}, E., \& {Renzini}, A. 2006, \aap, 453, L29

\bibitem[{{Cole} {et~al.}(2000){Cole}, {Lacey}, {Baugh}, \& {Frenk}}]{Cole00}
{Cole}, S., {Lacey}, C.~G., {Baugh}, C.~M., \& {Frenk}, C.~S. 2000, \mnras,
  319, 168

\bibitem[{{Combes}(2000)}]{Combes00}
{Combes}, F. 2000, in Building Galaxies; from the Primordial Universe to the
  Present, ed. {F.~Hammer, T.~X.~Thuan, V.~Cayatte, B.~Guiderdoni, \&
  J.~T.~Thanh Van }, 413

\bibitem[{{Conroy} {et~al.}(2007){Conroy}, {Ho}, \& {White}}]{Conroy07}
{Conroy}, C., {Ho}, S., \& {White}, M. 2007, \mnras, 379, 1491

\bibitem[{{Cool} {et~al.}(2008){Cool}, {Eisenstein}, {Fan}, {Fukugita},
  {Jiang}, {Maraston}, {Meiksin}, {Schneider}, \& {Wake}}]{Cool08}
{Cool}, R.~J., {et~al.} 2008, \apj, 682, 919

\bibitem[{{Croton} {et~al.}(2006){Croton}, {Springel}, {White}, {De Lucia},
  {Frenk}, {Gao}, {Jenkins}, {Kauffmann}, {Navarro}, \& {Yoshida}}]{Croton06}
{Croton}, D.~J., {et~al.} 2006, \mnras, 365, 11

\bibitem[{{De Lucia} \& {Blaizot}(2007)}]{DeLucia07}
{De Lucia}, G., \& {Blaizot}, J. 2007, \mnras, 375, 2

\bibitem[{{De Lucia} {et~al.}(2011){De Lucia}, {Fontanot}, {Wilman}, \&
  {Monaco}}]{DeLucia11}
{De Lucia}, G., {Fontanot}, F., {Wilman}, D., \& {Monaco}, P. 2011, \mnras,
  414, 1439

\bibitem[{{Efstathiou} {et~al.}(1982){Efstathiou}, {Lake}, \&
  {Negroponte}}]{Efstathiou82}
{Efstathiou}, G., {Lake}, G., \& {Negroponte}, J. 1982, \mnras, 199, 1069

\bibitem[{{Ellis} {et~al.}(1997){Ellis}, {Smail}, {Dressler}, {Couch},
  {Oemler}, {Butcher}, \& {Sharples}}]{Ellis97}
{Ellis}, R.~S., {Smail}, I., {Dressler}, A., {Couch}, W.~J., {Oemler}, A.~J.,
  {Butcher}, H., \& {Sharples}, R.~M. 1997, \apj, 483, 582

\bibitem[{{Erb} {et~al.}(2006){Erb}, {Shapley}, {Pettini}, {Steidel}, {Reddy},
  \& {Adelberger}}]{Erb06}
{Erb}, D.~K., {Shapley}, A.~E., {Pettini}, M., {Steidel}, C.~C., {Reddy},
  N.~A., \& {Adelberger}, K.~L. 2006, \apj, 644, 813

\bibitem[{{Faber} {et~al.}(2007){Faber}, {Willmer}, {Wolf}, {Koo}, {Weiner},
  {Newman}, {Im}, {Coil}, {Conroy}, {Cooper}, {Davis}, {Finkbeiner}, {Gerke},
  {Gebhardt}, {Groth}, {Guhathakurta}, {Harker}, {Kaiser}, {Kassin},
  {Kleinheinrich}, {Konidaris}, {Kron}, {Lin}, {Luppino}, {Madgwick},
  {Meisenheimer}, {Noeske}, {Phillips}, {Sarajedini}, {Schiavon}, {Simard},
  {Szalay}, {Vogt}, \& {Yan}}]{Faber07}
{Faber}, S.~M., {et~al.} 2007, \apj, 665, 265

\bibitem[{{Ferreras} {et~al.}(2009){Ferreras}, {Lisker}, {Pasquali},
  {Khochfar}, \& {Kaviraj}}]{Ferreras09}
{Ferreras}, I., {Lisker}, T., {Pasquali}, A., {Khochfar}, S., \& {Kaviraj}, S.
  2009, \mnras, 396, 1573

\bibitem[{{Fontanot} {et~al.}(2009){Fontanot}, {De Lucia}, {Monaco},
  {Somerville}, \& {Santini}}]{Fontanot09}
{Fontanot}, F., {De Lucia}, G., {Monaco}, P., {Somerville}, R.~S., \&
  {Santini}, P. 2009, \mnras, 397, 1776

\bibitem[{{Gallazzi} \& {Bell}(2009)}]{Gallazzi09b}
{Gallazzi}, A., \& {Bell}, E.~F. 2009, \apjs, 185, 253

\bibitem[{{Gallazzi} {et~al.}(2006){Gallazzi}, {Charlot}, {Brinchmann}, \&
  {White}}]{Gallazzi06}
{Gallazzi}, A., {Charlot}, S., {Brinchmann}, J., \& {White}, S.~D.~M. 2006,
  \mnras, 370, 1106

\bibitem[{{Gallazzi} {et~al.}(2005){Gallazzi}, {Charlot}, {Brinchmann},
  {White}, \& {Tremonti}}]{Gallazzi05}
{Gallazzi}, A., {Charlot}, S., {Brinchmann}, J., {White}, S.~D.~M., \&
  {Tremonti}, C.~A. 2005, \mnras, 362, 41

\bibitem[{{Gebhardt} {et~al.}(2003){Gebhardt}, {Faber}, {Koo}, {Im}, {Simard},
  {Illingworth}, {Phillips}, {Sarajedini}, {Vogt}, {Weiner}, \&
  {Willmer}}]{Gebhardt03}
{Gebhardt}, K., {et~al.} 2003, \apj, 597, 239

\bibitem[{{Harker} {et~al.}(2006){Harker}, {Schiavon}, {Weiner}, \&
  {Faber}}]{Harker06}
{Harker}, J.~J., {Schiavon}, R.~P., {Weiner}, B.~J., \& {Faber}, S.~M. 2006,
  \apjl, 647, L103

\bibitem[{{Holden} {et~al.}(2010){Holden}, {van der Wel}, {Kelson}, {Franx}, \&
  {Illingworth}}]{Holden10}
{Holden}, B.~P., {van der Wel}, A., {Kelson}, D.~D., {Franx}, M., \&
  {Illingworth}, G.~D. 2010, \apj, 724, 714

\bibitem[{{Holden} {et~al.}(2005){Holden}, {van der Wel}, {Franx},
  {Illingworth}, {Blakeslee}, {van Dokkum}, {Ford}, {Magee}, {Postman}, {Rix},
  \& {Rosati}}]{Holden05}
{Holden}, B.~P., {et~al.} 2005, \apjl, 620, L83

\bibitem[{{Hopkins} {et~al.}(2010){Hopkins}, {Bundy}, {Hernquist}, {Wuyts}, \&
  {Cox}}]{Hopkins10c}
{Hopkins}, P.~F., {Bundy}, K., {Hernquist}, L., {Wuyts}, S., \& {Cox}, T.~J.
  2010, \mnras, 401, 1099

\bibitem[{{Hopkins} {et~al.}(2009{\natexlab{a}}){Hopkins}, {Bundy}, {Murray},
  {Quataert}, {Lauer}, \& {Ma}}]{Hopkins09c}
{Hopkins}, P.~F., {Bundy}, K., {Murray}, N., {Quataert}, E., {Lauer}, T.~R., \&
  {Ma}, C.-P. 2009{\natexlab{a}}, \mnras, 398, 898

\bibitem[{{Hopkins} {et~al.}(2008{\natexlab{a}}){Hopkins}, {Cox}, {Kere{\v s}},
  \& {Hernquist}}]{Hopkins08b}
{Hopkins}, P.~F., {Cox}, T.~J., {Kere{\v s}}, D., \& {Hernquist}, L.
  2008{\natexlab{a}}, \apjs, 175, 390

\bibitem[{{Hopkins} {et~al.}(2009{\natexlab{b}}){Hopkins}, {Cox}, {Younger}, \&
  {Hernquist}}]{Hopkins09a}
{Hopkins}, P.~F., {Cox}, T.~J., {Younger}, J.~D., \& {Hernquist}, L.
  2009{\natexlab{b}}, \apj, 691, 1168

\bibitem[{{Hopkins} {et~al.}(2006{\natexlab{a}}){Hopkins}, {Hernquist}, {Cox},
  {Di Matteo}, {Robertson}, \& {Springel}}]{Hopkins06}
{Hopkins}, P.~F., {Hernquist}, L., {Cox}, T.~J., {Di Matteo}, T., {Robertson},
  B., \& {Springel}, V. 2006{\natexlab{a}}, \apjs, 163, 1

\bibitem[{{Hopkins} {et~al.}(2008{\natexlab{b}}){Hopkins}, {Hernquist}, {Cox},
  \& {Kere\v{s}}}]{Hopkins08a}
{Hopkins}, P.~F., {Hernquist}, L., {Cox}, T.~J., \& {Kere\v{s}}, D.
  2008{\natexlab{b}}, \apjs, 175, 356

\bibitem[{{Hopkins} {et~al.}(2006{\natexlab{b}}){Hopkins}, {Hernquist}, {Cox},
  {Robertson}, \& {Springel}}]{Hopkins06b}
{Hopkins}, P.~F., {Hernquist}, L., {Cox}, T.~J., {Robertson}, B., \&
  {Springel}, V. 2006{\natexlab{b}}, \apjs, 163, 50

\bibitem[{{Hopkins} {et~al.}(2009{\natexlab{c}}){Hopkins}, {Somerville}, {Cox},
  {Hernquist}, {Jogee}, {Kere{\v s}}, {Ma}, {Robertson}, \&
  {Stewart}}]{Hopkins09b}
{Hopkins}, P.~F., {et~al.} 2009{\natexlab{c}}, \mnras, 397, 802

\bibitem[{{Ilbert} {et~al.}(2010){Ilbert}, {Salvato}, {Le Floc'h}, {Aussel},
  {Capak}, {McCracken}, {Mobasher}, {Kartaltepe}, {Scoville}, {Sanders},
  {Arnouts}, {Bundy}, {Cassata}, {Kneib}, {Koekemoer}, {Le F{\`e}vre}, {Lilly},
  {Surace}, {Taniguchi}, {Tasca}, {Thompson}, {Tresse}, {Zamojski}, {Zamorani},
  \& {Zucca}}]{Ilbert10}
{Ilbert}, O., {et~al.} 2010, \apj, 709, 644

\bibitem[{{Jogee} {et~al.}(2009){Jogee}, {Miller}, {Penner}, {Skelton},
  {Conselice}, {Somerville}, {Bell}, {Zheng}, {Rix}, {Robaina}, {Barazza},
  {Barden}, {Borch}, {Beckwith}, {Caldwell}, {Peng}, {Heymans}, {McIntosh},
  {H{\"a}u{\ss}ler}, {Jahnke}, {Meisenheimer}, {Sanchez}, {Wisotzki}, {Wolf},
  \& {Papovich}}]{Jogee09}
{Jogee}, S., {et~al.} 2009, \apj, 697, 1971

\bibitem[{{Kauffmann} {et~al.}(2003){Kauffmann}, {Heckman}, {White}, {Charlot},
  {Tremonti}, {Brinchmann}, {Bruzual}, {Peng}, {Seibert}, {Bernardi},
  {Blanton}, {Brinkmann}, {Castander}, {Cs{\'a}bai}, {Fukugita}, {Ivezic},
  {Munn}, {Nichol}, {Padmanabhan}, {Thakar}, {Weinberg}, \&
  {York}}]{Kauffmann03a}
{Kauffmann}, G., {et~al.} 2003, \mnras, 341, 33

\bibitem[{{Kewley} \& {Ellison}(2008)}]{Kewley08}
{Kewley}, L.~J., \& {Ellison}, S.~L. 2008, \apj, 681, 1183

\bibitem[{{Kodama} \& {Arimoto}(1997)}]{Kodama97}
{Kodama}, T., \& {Arimoto}, N. 1997, \aap, 320, 41

\bibitem[{{Kriek} {et~al.}(2008){Kriek}, {van der Wel}, {van Dokkum}, {Franx},
  \& {Illingworth}}]{Kriek08}
{Kriek}, M., {van der Wel}, A., {van Dokkum}, P.~G., {Franx}, M., \&
  {Illingworth}, G.~D. 2008, \apj, 682, 896

\bibitem[{{Labb{\'e}} {et~al.}(2005){Labb{\'e}}, {Huang}, {Franx}, {Rudnick},
  {Barmby}, {Daddi}, {van Dokkum}, {Fazio}, {Schreiber}, {Moorwood}, {Rix},
  {R{\"o}ttgering}, {Trujillo}, \& {van der Werf}}]{Labbe05}
{Labb{\'e}}, I., {et~al.} 2005, \apjl, 624, L81

\bibitem[{{Lauer} {et~al.}(2007){Lauer}, {Faber}, {Richstone}, {Gebhardt},
  {Tremaine}, {Postman}, {Dressler}, {Aller}, {Filippenko}, {Green}, {Ho},
  {Kormendy}, {Magorrian}, \& {Pinkney}}]{Lauer07}
{Lauer}, T.~R., {et~al.} 2007, \apj, 662, 808

\bibitem[{{Lee} {et~al.}(2010){Lee}, {Ferguson}, {Somerville}, {Wiklind}, \&
  {Giavalisco}}]{Lee10}
{Lee}, S.-K., {Ferguson}, H.~C., {Somerville}, R.~S., {Wiklind}, T., \&
  {Giavalisco}, M. 2010, \apj, 725, 1644

\bibitem[{{Lin} {et~al.}(2008){Lin}, {Patton}, {Koo}, {Casteels}, {Conselice},
  {Faber}, {Lotz}, {Willmer}, {Hsieh}, {Chiueh}, {Newman}, {Novak}, {Weiner},
  \& {Cooper}}]{Lin08}
{Lin}, L., {et~al.} 2008, \apj, 681, 232

\bibitem[{{Lotz} {et~al.}(2011){Lotz}, {Jonsson}, {Cox}, {Croton}, {Primack},
  {Somerville}, \& {Stewart}}]{Lotz11}
{Lotz}, J.~M., {Jonsson}, P., {Cox}, T.~J., {Croton}, D., {Primack}, J.~R.,
  {Somerville}, R.~S., \& {Stewart}, K. 2011, \apj, 742, 103

\bibitem[{{Lotz} {et~al.}(2010{\natexlab{a}}){Lotz}, {Jonsson}, {Cox}, \&
  {Primack}}]{Lotz10a}
{Lotz}, J.~M., {Jonsson}, P., {Cox}, T.~J., \& {Primack}, J.~R.
  2010{\natexlab{a}}, \mnras, 404, 590

\bibitem[{{Lotz} {et~al.}(2010{\natexlab{b}}){Lotz}, {Jonsson}, {Cox}, \&
  {Primack}}]{Lotz10b}
---. 2010{\natexlab{b}}, \mnras, 404, 575

\bibitem[{{Lotz} {et~al.}(2006){Lotz}, {Madau}, {Giavalisco}, {Primack}, \&
  {Ferguson}}]{Lotz06}
{Lotz}, J.~M., {Madau}, P., {Giavalisco}, M., {Primack}, J., \& {Ferguson},
  H.~C. 2006, \apj, 636, 592

\bibitem[{{Madgwick} {et~al.}(2002){Madgwick}, {Lahav}, {Baldry}, {Baugh},
  {Bland-Hawthorn}, {Bridges}, {Cannon}, {Cole}, {Colless}, {Collins}, {Couch},
  {Dalton}, {De Propris}, {Driver}, {Efstathiou}, {Ellis}, {Frenk},
  {Glazebrook}, {Jackson}, {Lewis}, {Lumsden}, {Maddox}, {Norberg}, {Peacock},
  {Peterson}, {Sutherland}, \& {Taylor}}]{Madgwick02}
{Madgwick}, D.~S., {et~al.} 2002, \mnras, 333, 133

\bibitem[{{Man} {et~al.}(2012){Man}, {Toft}, {Zirm}, {Wuyts}, \& {van der
  Wel}}]{Man11}
{Man}, A.~W.~S., {Toft}, S., {Zirm}, A.~W., {Wuyts}, S., \& {van der Wel}, A.
  2012, \apj, 744, 85

\bibitem[{{Maraston}(2005)}]{Maraston05}
{Maraston}, C. 2005, \mnras, 362, 799

\bibitem[{{Masjedi} {et~al.}(2006){Masjedi}, {Hogg}, {Cool}, {Eisenstein},
  {Blanton}, {Zehavi}, {Berlind}, {Bell}, {Schneider}, {Warren}, \&
  {Brinkmann}}]{Masjedi06}
{Masjedi}, M., {et~al.} 2006, \apj, 644, 54

\bibitem[{{Matsuoka} \& {Kawara}(2010)}]{Matsuoka10}
{Matsuoka}, Y., \& {Kawara}, K. 2010, \mnras, 405, 100

\bibitem[{{McIntosh} {et~al.}(2008){McIntosh}, {Guo}, {Hertzberg}, {Katz},
  {Mo}, {van den Bosch}, \& {Yang}}]{McIntosh08}
{McIntosh}, D.~H., {Guo}, Y., {Hertzberg}, J., {Katz}, N., {Mo}, H.~J., {van
  den Bosch}, F.~C., \& {Yang}, X. 2008, \mnras, 388, 1537

\bibitem[{{McIntosh} {et~al.}(2005){McIntosh}, {Zabludoff}, {Rix}, \&
  {Caldwell}}]{McIntosh05b}
{McIntosh}, D.~H., {Zabludoff}, A.~I., {Rix}, H., \& {Caldwell}, N. 2005, \apj,
  619, 193

\bibitem[{{Mo} {et~al.}(1998){Mo}, {Mao}, \& {White}}]{Mo98}
{Mo}, H.~J., {Mao}, S., \& {White}, S.~D.~M. 1998, \mnras, 295, 319

\bibitem[{{Monaco} {et~al.}(2007){Monaco}, {Fontanot}, \& {Taffoni}}]{Monaco07}
{Monaco}, P., {Fontanot}, F., \& {Taffoni}, G. 2007, \mnras, 375, 1189

\bibitem[{{Moustakas} {et~al.}(2011){Moustakas}, {Zaritsky}, {Brown}, {Cool},
  {Dey}, {Eisenstein}, {Gonzalez}, {Jannuzi}, {Jones}, {Kochanek}, {Murray}, \&
  {Wild}}]{Moustakas11}
{Moustakas}, J., {et~al.} 2011, arXiv:1112.3300

\bibitem[{{Naab} {et~al.}(2009){Naab}, {Johansson}, \& {Ostriker}}]{Naab09}
{Naab}, T., {Johansson}, P.~H., \& {Ostriker}, J.~P. 2009, \apjl, 699, L178

\bibitem[{{Navarro} {et~al.}(1997){Navarro}, {Frenk}, \& {White}}]{Navarro97}
{Navarro}, J.~F., {Frenk}, C.~S., \& {White}, S.~D.~M. 1997, \apj, 490, 493

\bibitem[{{Padilla} {et~al.}(2011){Padilla}, {Christlein}, {Gawiser}, \&
  {Marchesini}}]{Padilla11}
{Padilla}, N., {Christlein}, D., {Gawiser}, E., \& {Marchesini}, D. 2011, \aap,
  531, A142

\bibitem[{{Papovich} {et~al.}(2011){Papovich}, {Finkelstein}, {Ferguson},
  {Lotz}, \& {Giavalisco}}]{Papovich11}
{Papovich}, C., {Finkelstein}, S.~L., {Ferguson}, H.~C., {Lotz}, J.~M., \&
  {Giavalisco}, M. 2011, \mnras, 412, 1123

\bibitem[{{Robaina} {et~al.}(2010){Robaina}, {Bell}, {van der Wel},
  {Somerville}, {Skelton}, {McIntosh}, {Meisenheimer}, \& {Wolf}}]{Robaina10}
{Robaina}, A.~R., {Bell}, E.~F., {van der Wel}, A., {Somerville}, R.~S.,
  {Skelton}, R.~E., {McIntosh}, D.~H., {Meisenheimer}, K., \& {Wolf}, C. 2010,
  \apj, 719, 844

\bibitem[{{Robaina} {et~al.}(2009){Robaina}, {Bell}, {Skelton}, {Mc Intosh},
  {Somerville}, {Zheng}, {Rix}, {Bacon}, {Balogh}, {Barazza}, {Barden},
  {B{\"o}hm}, {Caldwell}, {Gallazzi}, {Gray}, {H{\"a}ussler}, {Heymans},
  {Jahnke}, {Jogee}, {van Kampen}, {Lane}, {Meisenheimer}, {Papovich}, {Peng},
  {S{\'a}nchez}, {Skibba}, {Taylor}, {Wisotzki}, \& {Wolf}}]{Robaina09a}
{Robaina}, A.~R., {et~al.} 2009, \apj, 704, 324

\bibitem[{{Ruhland} {et~al.}(2009){Ruhland}, {Bell}, {H{\"a}u{\ss}ler},
  {Taylor}, {Barden}, \& {McIntosh}}]{Ruhland09}
{Ruhland}, C., {Bell}, E.~F., {H{\"a}u{\ss}ler}, B., {Taylor}, E.~N., {Barden},
  M., \& {McIntosh}, D.~H. 2009, \apj, 695, 1058

\bibitem[{{Rusin} \& {Kochanek}(2005)}]{Rusin05}
{Rusin}, D., \& {Kochanek}, C.~S. 2005, \apj, 623, 666

\bibitem[{{Savaglio} {et~al.}(2005){Savaglio}, {Glazebrook}, {Le Borgne},
  {Juneau}, {Abraham}, {Chen}, {Crampton}, {McCarthy}, {Carlberg}, {Marzke},
  {Roth}, {J{\o}rgensen}, \& {Murowinski}}]{Savaglio05}
{Savaglio}, S., {et~al.} 2005, \apj, 635, 260

\bibitem[{{Schweizer} \& {Seitzer}(1992)}]{Schweizer92}
{Schweizer}, F., \& {Seitzer}, P. 1992, \aj, 104, 1039

\bibitem[{{Skelton} {et~al.}(2009){Skelton}, {Bell}, \&
  {Somerville}}]{Skelton09a}
{Skelton}, R.~E., {Bell}, E.~F., \& {Somerville}, R.~S. 2009, \apjl, 699, L9

\bibitem[{{Somerville} {et~al.}(2008){Somerville}, {Hopkins}, {Cox},
  {Robertson}, \& {Hernquist}}]{Somerville08}
{Somerville}, R.~S., {Hopkins}, P.~F., {Cox}, T.~J., {Robertson}, B.~E., \&
  {Hernquist}, L. 2008, \mnras, 391, 481

\bibitem[{{Somerville} \& {Kolatt}(1999)}]{SK99}
{Somerville}, R.~S., \& {Kolatt}, T.~S. 1999, \mnras, 305, 1

\bibitem[{{Somerville} {et~al.}(2001){Somerville}, {Primack}, \&
  {Faber}}]{Somerville01}
{Somerville}, R.~S., {Primack}, J.~R., \& {Faber}, S.~M. 2001, \mnras, 320, 504

\bibitem[{{Stewart} {et~al.}(2009){Stewart}, {Bullock}, {Wechsler}, \&
  {Maller}}]{Stewart09b}
{Stewart}, K.~R., {Bullock}, J.~S., {Wechsler}, R.~H., \& {Maller}, A.~H. 2009,
  \apj, 702, 307

\bibitem[{{Strateva} {et~al.}(2001){Strateva}, {Ivezi{\'c}}, {Knapp},
  {Narayanan}, {Strauss}, {Gunn}, {Lupton}, {Schlegel}, {Bahcall}, {Brinkmann},
  {Brunner}, {Budav{\'a}ri}, {Csabai}, {Castander}, {Doi}, {Fukugita},
  {Gy\H{o}ry}, {Hamabe}, {Hennessy}, {Ichikawa}, {Kunszt}, {Lamb}, {McKay},
  {Okamura}, {Racusin}, {Sekiguchi}, {Schneider}, {Shimasaku}, \&
  {York}}]{Strateva01}
{Strateva}, I., {et~al.} 2001, \aj, 122, 1861

\bibitem[{{Tal} {et~al.}(2009){Tal}, {van Dokkum}, {Nelan}, \&
  {Bezanson}}]{Tal09}
{Tal}, T., {van Dokkum}, P.~G., {Nelan}, J., \& {Bezanson}, R. 2009, \aj, 138,
  1417

\bibitem[{{Tinsley}(1980)}]{Tinsley80}
{Tinsley}, B.~M. 1980, \fcp, 5, 287

\bibitem[{{Tojeiro} \& {Percival}(2010)}]{Tojeiro10}
{Tojeiro}, R., \& {Percival}, W.~J. 2010, \mnras, 405, 2534

\bibitem[{{Tojeiro} \& {Percival}(2011)}]{Tojeiro11b}
---. 2011, \mnras, 417, 1114

\bibitem[{{Toomre}(1977)}]{Toomre77}
{Toomre}, A. 1977, in Evolution of Galaxies and Stellar Populations, ed. B.~M.
  {Tinsley} \& R.~B. {Larson}, 401

\bibitem[{{Tremonti} {et~al.}(2004){Tremonti}, {Heckman}, {Kauffmann},
  {Brinchmann}, {Charlot}, {White}, {Seibert}, {Peng}, {Schlegel}, {Uomoto},
  {Fukugita}, \& {Brinkmann}}]{Tremonti04}
{Tremonti}, C.~A., {et~al.} 2004, \apj, 613, 898

\bibitem[{{Treu} {et~al.}(2005){Treu}, {Ellis}, {Liao}, {van Dokkum}, {Tozzi},
  {Coil}, {Newman}, {Cooper}, \& {Davis}}]{Treu05}
{Treu}, T., {et~al.} 2005, \apj, 633, 174

\bibitem[{{van der Wel} {et~al.}(2009{\natexlab{a}}){van der Wel}, {Bell}, {van
  den Bosch}, {Gallazzi}, \& {Rix}}]{vanderWel09b}
{van der Wel}, A., {Bell}, E.~F., {van den Bosch}, F.~C., {Gallazzi}, A., \&
  {Rix}, H. 2009{\natexlab{a}}, \apj, 698, 1232

\bibitem[{{van der Wel} {et~al.}(2005){van der Wel}, {Franx}, {van Dokkum},
  {Rix}, {Illingworth}, \& {Rosati}}]{vanderWel05}
{van der Wel}, A., {Franx}, M., {van Dokkum}, P.~G., {Rix}, H., {Illingworth},
  G.~D., \& {Rosati}, P. 2005, \apj, 631, 145

\bibitem[{{van der Wel} {et~al.}(2004){van der Wel}, {Franx}, {van Dokkum}, \&
  {Rix}}]{vanderWel04}
{van der Wel}, A., {Franx}, M., {van Dokkum}, P.~G., \& {Rix}, H.-W. 2004,
  \apjl, 601, L5

\bibitem[{{van der Wel} {et~al.}(2006){van der Wel}, {Franx}, {Wuyts}, {van
  Dokkum}, {Huang}, {Rix}, \& {Illingworth}}]{vanderWel06}
{van der Wel}, A., {Franx}, M., {Wuyts}, S., {van Dokkum}, P.~G., {Huang}, J.,
  {Rix}, H.-W., \& {Illingworth}, G.~D. 2006, \apj, 652, 97

\bibitem[{{van der Wel} {et~al.}(2008){van der Wel}, {Holden}, {Zirm}, {Franx},
  {Rettura}, {Illingworth}, \& {Ford}}]{vanderWel08}
{van der Wel}, A., {Holden}, B.~P., {Zirm}, A.~W., {Franx}, M., {Rettura}, A.,
  {Illingworth}, G.~D., \& {Ford}, H.~C. 2008, \apj, 688, 48

\bibitem[{{van der Wel} {et~al.}(2009{\natexlab{b}}){van der Wel}, {Rix},
  {Holden}, {Bell}, \& {Robaina}}]{vanderWel09}
{van der Wel}, A., {Rix}, H., {Holden}, B.~P., {Bell}, E.~F., \& {Robaina},
  A.~R. 2009{\natexlab{b}}, \apjl, 706, L120

\bibitem[{{van Dokkum}(2005)}]{vanDokkum05}
{van Dokkum}, P.~G. 2005, \aj, 130, 2647

\bibitem[{{van Dokkum}(2008)}]{vanDokkum08}
---. 2008, \apj, 674, 29

\bibitem[{{van Dokkum} \& {Ellis}(2003)}]{vanDokkumEllis03}
{van Dokkum}, P.~G., \& {Ellis}, R.~S. 2003, \apjl, 592, L53

\bibitem[{{van Dokkum} \& {Franx}(1996)}]{vanDokkumFranx96}
{van Dokkum}, P.~G., \& {Franx}, M. 1996, \mnras, 281, 985

\bibitem[{{van Dokkum} \& {Franx}(2001)}]{vanDokkumFranx01}
---. 2001, \apj, 553, 90

\bibitem[{{van Dokkum} {et~al.}(2000){van Dokkum}, {Franx}, {Fabricant},
  {Illingworth}, \& {Kelson}}]{vanDokkum00}
{van Dokkum}, P.~G., {Franx}, M., {Fabricant}, D., {Illingworth}, G.~D., \&
  {Kelson}, D.~D. 2000, \apj, 541, 95

\bibitem[{{van Dokkum} {et~al.}(1998{\natexlab{a}}){van Dokkum}, {Franx},
  {Kelson}, \& {Illingworth}}]{vanDokkum98}
{van Dokkum}, P.~G., {Franx}, M., {Kelson}, D.~D., \& {Illingworth}, G.~D.
  1998{\natexlab{a}}, \apjl, 504, L17+

\bibitem[{{van Dokkum} {et~al.}(1998{\natexlab{b}}){van Dokkum}, {Franx},
  {Kelson}, {Illingworth}, {Fisher}, \& {Fabricant}}]{vanDokkum98b}
{van Dokkum}, P.~G., {Franx}, M., {Kelson}, D.~D., {Illingworth}, G.~D.,
  {Fisher}, D., \& {Fabricant}, D. 1998{\natexlab{b}}, \apj, 500, 714

\bibitem[{{van Dokkum} \& {Stanford}(2003)}]{vanDokkumStanford03}
{van Dokkum}, P.~G., \& {Stanford}, S.~A. 2003, \apj, 585, 78

\bibitem[{{van Dokkum} \& {van der Marel}(2007)}]{vDvdMarel07}
{van Dokkum}, P.~G., \& {van der Marel}, R.~P. 2007, \apj, 655, 30

\bibitem[{{Wake} {et~al.}(2006){Wake}, {Nichol}, {Eisenstein}, {Loveday},
  {Edge}, {Cannon}, {Smail}, {Schneider}, {Scranton}, {Carson}, {Ross},
  {Brunner}, {Colless}, {Couch}, {Croom}, {Driver}, {da {\^A}ngela}, {Jester},
  {de Propris}, {Drinkwater}, {Bland-Hawthorn}, {Pimbblet}, {Roseboom},
  {Shanks}, {Sharp}, \& {Brinkmann}}]{Wake06}
{Wake}, D.~A., {et~al.} 2006, \mnras, 372, 537

\bibitem[{{Wake} {et~al.}(2008){Wake}, {Sheth}, {Nichol}, {Baugh},
  {Bland-Hawthorn}, {Colless}, {Couch}, {Croom}, {de Propris}, {Drinkwater},
  {Edge}, {Loveday}, {Lam}, {Pimbblet}, {Roseboom}, {Ross}, {Schneider},
  {Shanks}, \& {Sharp}}]{Wake08}
---. 2008, \mnras, 387, 1045

\bibitem[{{Whitaker} {et~al.}(2010){Whitaker}, {van Dokkum}, {Brammer},
  {Kriek}, {Franx}, {Labb{\'e}}, {Marchesini}, {Quadri}, {Bezanson},
  {Illingworth}, {Lee}, {Muzzin}, {Rudnick}, \& {Wake}}]{Whitaker10}
{Whitaker}, K.~E., {et~al.} 2010, \apj, 719, 1715

\bibitem[{{Whitaker} {et~al.}(2011){Whitaker}, {Labb{\'e}}, {van Dokkum},
  {Brammer}, {Kriek}, {Marchesini}, {Quadri}, {Franx}, {Muzzin}, {Williams},
  {Bezanson}, {Illingworth}, {Lee}, {Lundgren}, {Nelson}, {Rudnick}, {Tal}, \&
  {Wake}}]{Whitaker11}
---. 2011, \apj, 735, 86

\bibitem[{{White} {et~al.}(2007){White}, {Zheng}, {Brown}, {Dey}, \&
  {Jannuzi}}]{White07}
{White}, M., {Zheng}, Z., {Brown}, M.~J.~I., {Dey}, A., \& {Jannuzi}, B.~T.
  2007, \apjl, 655, L69

\bibitem[{{Williams} {et~al.}(2011){Williams}, {Quadri}, \&
  {Franx}}]{Williams11}
{Williams}, R.~J., {Quadri}, R.~F., \& {Franx}, M. 2011, \apjl, 738, L25

\bibitem[{{Williams} {et~al.}(2009){Williams}, {Quadri}, {Franx}, {van Dokkum},
  \& {Labb{\'e}}}]{Williams09}
{Williams}, R.~J., {Quadri}, R.~F., {Franx}, M., {van Dokkum}, P., \&
  {Labb{\'e}}, I. 2009, \apj, 691, 1879

\bibitem[{{Wuyts} {et~al.}(2004){Wuyts}, {van Dokkum}, {Kelson}, {Franx}, \&
  {Illingworth}}]{Wuyts04}
{Wuyts}, S., {van Dokkum}, P.~G., {Kelson}, D.~D., {Franx}, M., \&
  {Illingworth}, G.~D. 2004, \apj, 605, 677

\bibitem[{{Zahid} {et~al.}(2011){Zahid}, {Kewley}, \& {Bresolin}}]{Zahid11}
{Zahid}, H.~J., {Kewley}, L.~J., \& {Bresolin}, F. 2011, \apj, 730, 137

\end{thebibliography}

\end{document}